\documentclass[epj]{svjour}
\usepackage{graphics}
\usepackage{graphicx}
\usepackage{amsmath}
\usepackage{bm}
\usepackage{epsfig}
\usepackage{epstopdf} 
\usepackage{subfigure}

\usepackage{grffile}

\newcommand{\be}{\begin{equation}}
\newcommand{\ee}{\end{equation}}
\newcommand{\bea}{\begin{eqnarray}}
\newcommand{\eea}{\end{eqnarray}}
\newcommand{\bfk}{\mbox{\boldmath $k$}}
\def\kt{k_\perp}
\def\pt{p_\perp}
\newcommand{\bfp}{\mbox{\boldmath $p$}}
\newcommand{\bfq}{\mbox{\boldmath $q$}}
\newcommand{\bfP}{\mbox{\boldmath $P$}}
\newcommand{\bfS}{\mbox{\boldmath $S$}}
\newcommand{\pup}{p^\uparrow}
\newcommand{\qup}{q^\uparrow}

\begin{document}
\title{TMDs and SSAs in hadronic interactions}
\author{E.C.~Aschenauer\inst{1} \and U. D'Alesio\inst{2,3} \and F.~Murgia\inst{3
}                     
%
%
\institute{Physics Department, Brookhaven National Laboratory, Upton, New York 11973-5000, USA \and Dipartimento di Fisica, Universit\`a di Cagliari, Cittadella Universitaria, I-09042 Monserrato (CA), Italy \and INFN, Sezione di Cagliari, C.P.~170, I-09042 Monserrato (CA), Italy}}
\date{Received: date / Revised version: date}
%
\abstract{
We present an overview on the current experimental and phenomenological status of transverse single spin asymmetries (tSSAs) in proton-proton collisions. In particular, we focus on large-$p_T$ inclusive pion, photon, jet, pion-jet production and Drell-Yan processes. For all of them theoretical estimates are given in terms of a generalised parton model (GPM) based on a transverse momentum dependent (TMD) factorisation scheme. Comparisons with the corresponding results in a collinear twist-3 formalism and in a modified GPM approach are also made. On the experimental side, a selection of the most interesting and recent results from RHIC is presented.
} 
\maketitle

\section{Introduction}
\label{intro}

In recent years, transverse spin phenomena have gained substantial attention as they offer the unique opportunity to expand our current one-dimensional picture of the nucleon by imaging the proton in both longitudinal and transverse momentum and impact parameter space. At the same time these phenomena can help in understanding the basics of colour interactions in QCD and how they manifest themselves in different processes.

Understanding the large transverse Single-Spin Asymmetries (tSSAs), abundantly observed in high-$p_T$ inclusive hadron production in high-energy proton-proton experiments, is certainly one of the major challenges and fascinating issues in today's hadronic physics. In fact, because of helicity conservation (in the massless limit) intrinsic to QED and QCD interactions, large tSSAs cannot be generated in the hard elementary processes. Their persisting at large transverse momenta must therefore be related to non perturbative properties of the nucleon structure, such as parton intrinsic and orbital motion. An unambiguous understanding of the origin of tSSAs would then allow a deeper knowledge of the nucleon structure.

Two different, although somewhat related, approaches are thought to be the possible key solution to the problem~\cite{D'Alesio:2007jt}. One (known as ``twist-3 approach"), based on the well-known collinear QCD factorisation scheme and suitable for high-energy single-scale processes, involves as basic quantities higher-twist quark-gluon-quark correlations in the nucleon as well as in the hadronisation process . The second approach (named GPM in the following) is based on a more phenomenological generalisation of the parton model, with the inclusion, in the factorisation scheme, of transverse momentum dependent partonic distribution and fragmentation functions (TMDs).
See Refs.\cite{Efremov:1981sh,Efremov:1984ip,Qiu:1991pp,Qiu:1991wg,Qiu:1998ia,Ji:2006vf,Kouvaris:2006zy,Kanazawa:2014dca,Kanazawa:2014nea,Anselmino:1994tv,D'Alesio:2004up,Anselmino:2005sh,Anselmino:2012rq,Anselmino:2013rya} and references therein, for a more detailed account of the two approaches.

Factorisation in the TMD formalism has been proven for double energy scale (a hard and a soft one) processes like semi-inclusive deeply inelastic scattering (SIDIS), Drell-Yan (DY) and $e^+e^-$ annihilation processes~\cite{Collins:1984kg,Ji:2004xq,Collins:2011zzd,GarciaEchevarria:2011rb},  allowing the extraction of some relevant TMDs, like the Sivers~\cite{Sivers:1989cc,Sivers:1990fh} and the Collins~\cite{Collins:1992kk} functions.
Notice that twist-3 correlations are somehow related to transverse-momentum moments of corresponding TMDs.

\begin{figure*}[th!]
 \begin{center}
   \centering
    \includegraphics[width=1.90\columnwidth]{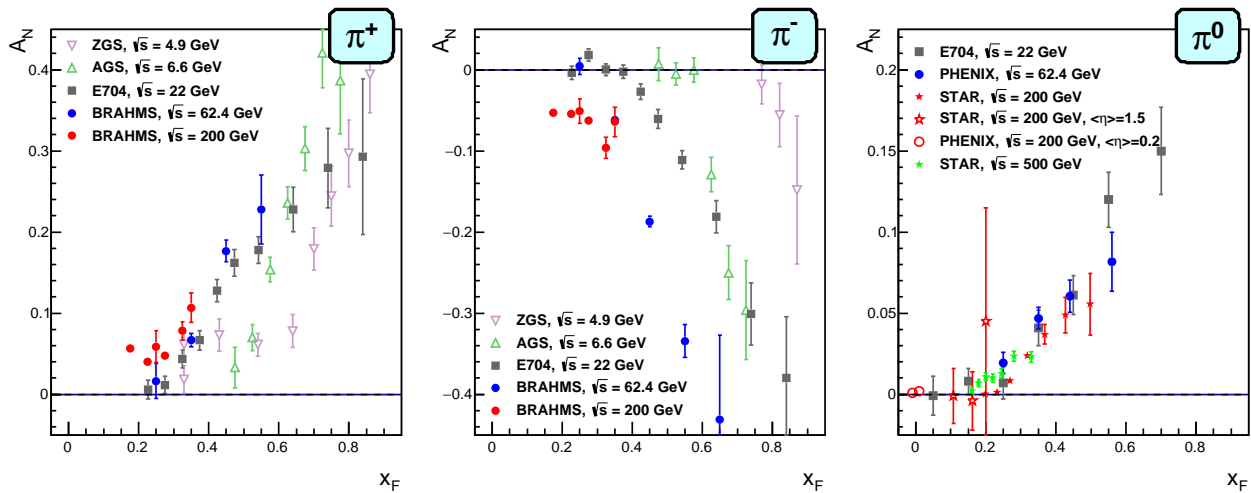}
     \caption{Transverse single spin asymmetry measurements for charged and neutral pions at
              different center-of-mass energies as a function of Feynman-$x$, $x_F$.}
  \label{AN.world}
 \end{center}
\end{figure*}

Concerning the experimental side, a myriad of new techniques and technologies made it possible to inaugurate the Relativistic Heavy Ion Collider (RHIC) at Brookhaven National Laboratory as the world's first high-energy polarised proton collider in December 2001. This unique environment provides opportunities to study the polarised quark and gluon spin structure of the proton and QCD dynamics at a high-energy scale, therefore it is complementary to existing SIDIS experiments. The polarised proton beam program at RHIC has in the past and will continue to address:
\begin{itemize}
\item How do quarks and gluons hadronise into final-state particles?
\item How do spin phenomena in QCD arise at the quark and gluon level?
\item How can we access the full 3D structure of the nucleon?
\end{itemize}

After a general overview on some of the most relevant experimental results we will focus on the phenomenological study of azimuthal and tSSAs in $pp$ collisions within the GPM approach: this indeed offers a powerful tool to describe many data sets for the inclusive cases and their main features, representing at the same time a window into possible factorisation breaking effects. We will also comment, wherever appropriate, on the corresponding results in the twist-3 approach. A general overview on TMDs and their phenomenology in SIDIS and $e^+e^-$ annihilation processes can be found in Ref.~\cite{Boglione:2015zyc} (this Special Issue).

For their relevance we will present and discuss in some detail a selection of results from RHIC, that has provided and is still providing with the most interesting and challenging experimental data. For a recent discussion of the potential role of SSA studies for the fixed-target experiment AFTER, proposed at the Large Hadron Collider (LHC), see Ref.~\cite{Anselmino:2015eoa}.
It is also worth to mention the proposed polarised target and beam program with SeaQuest~\cite{Brown:2014sea,Isenhower:2012vh} at FermiLab that will provide the possibility to study TMDs for sea and valence quarks through DY production.
Although there is no dedicated TMD experimental program, LHC would also offer a nice opportunity to investigate, at the largest available center-of-mass energy and transverse momentum, several TMD observables and effects involved in azimuthal asymmetries for unpolarised $pp$ and $pA$ collisions.

\section{Experimental Results}
\label{sec:1.exp.res}

Results from the PHENIX~\cite{Adcox:2003zm} and STAR~\cite{Ackermann:2002ad} Collaborations have shown that large transverse single spin asymmetries for inclusive hadron production, $A_N$, that were first seen in $pp$ collisions at fixed-target energies and modest $p_T$ (the transverse momentum of the final hadron), extend to the highest RHIC center-of-mass (c.m.) energies, $\sqrt{s}=500$ GeV and surprisingly large $p_T$. These asymmetries are defined as
$A_N = \frac{d\sigma^\uparrow - d\sigma^\downarrow}
           {d\sigma^\uparrow + d\sigma^\downarrow}$, where $\uparrow, \downarrow$ represent the two opposite spin
orientations perpendicular to the scattering plane (see Sec.~\ref{pph} for further details).
Figure~\ref{AN.world} summarizes the measured asymmetries from different experiments as a function of Feynman-$x$, $x_F =2p_L/\sqrt s \sim x_1-x_2$, where $p_L$ is the c.m.~longitudinal momentum of the final hadron and $x_{1,2}$ the initial parton light-cone momentum fractions. Surprisingly the asymmetries are nearly independent of $\sqrt{s}$ over a very wide range ($\sqrt{s}$: 4.9 GeV to 500 GeV).

To understand the underlying physics being responsible for the observed SSAs one has to go beyond the conventional collinear parton picture in the hard scattering. As already stated in the introduction, two theoretical formalisms have been proposed to generate sizeable SSAs in the QCD framework:
one based on transverse momentum dependent (TMD) parton distribution functions (PDFs) and fragmentation functions (FFs), and the other based on collinear twist-3 quark-gluon-quark correlations in the initial state proton or in the fragmentation process.
As the SSAs for inclusive hadrons cannot discriminate between these different approaches, nor among the different mechanisms within the same formalism (initial vs. final state effects), the focus has in the recent years shifted to observables that could help in disentangling them clarifying their effective role, and, at the same time, will be able to give new insight into the transverse spin structure of hadrons.

\begin{figure*}[ht!]
 \begin{center}
  \subfigure[]{
   \centering
    \includegraphics[width=0.90\columnwidth]{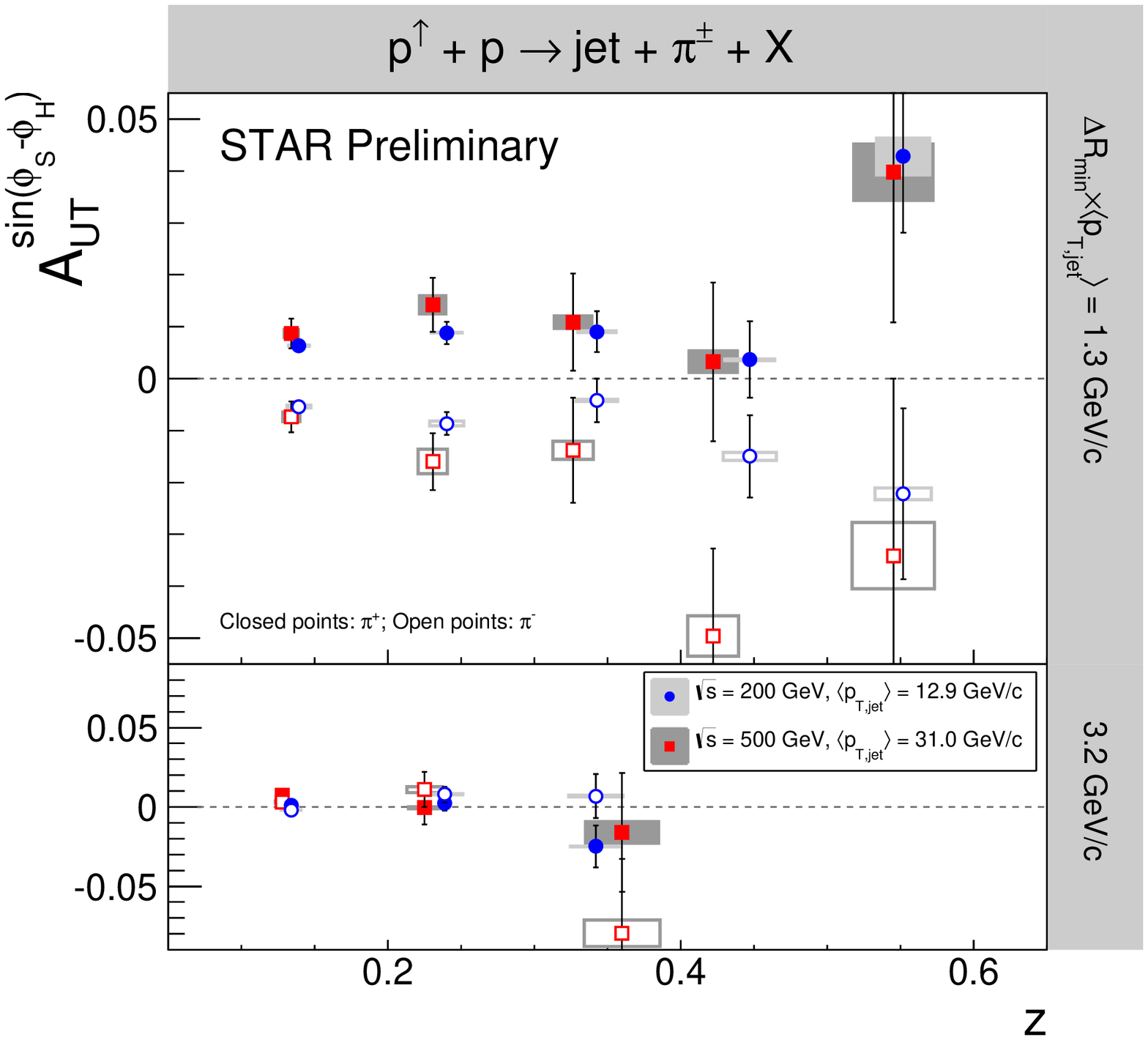} }
  \subfigure[]{
   \centering
    \includegraphics[width=0.90\columnwidth]{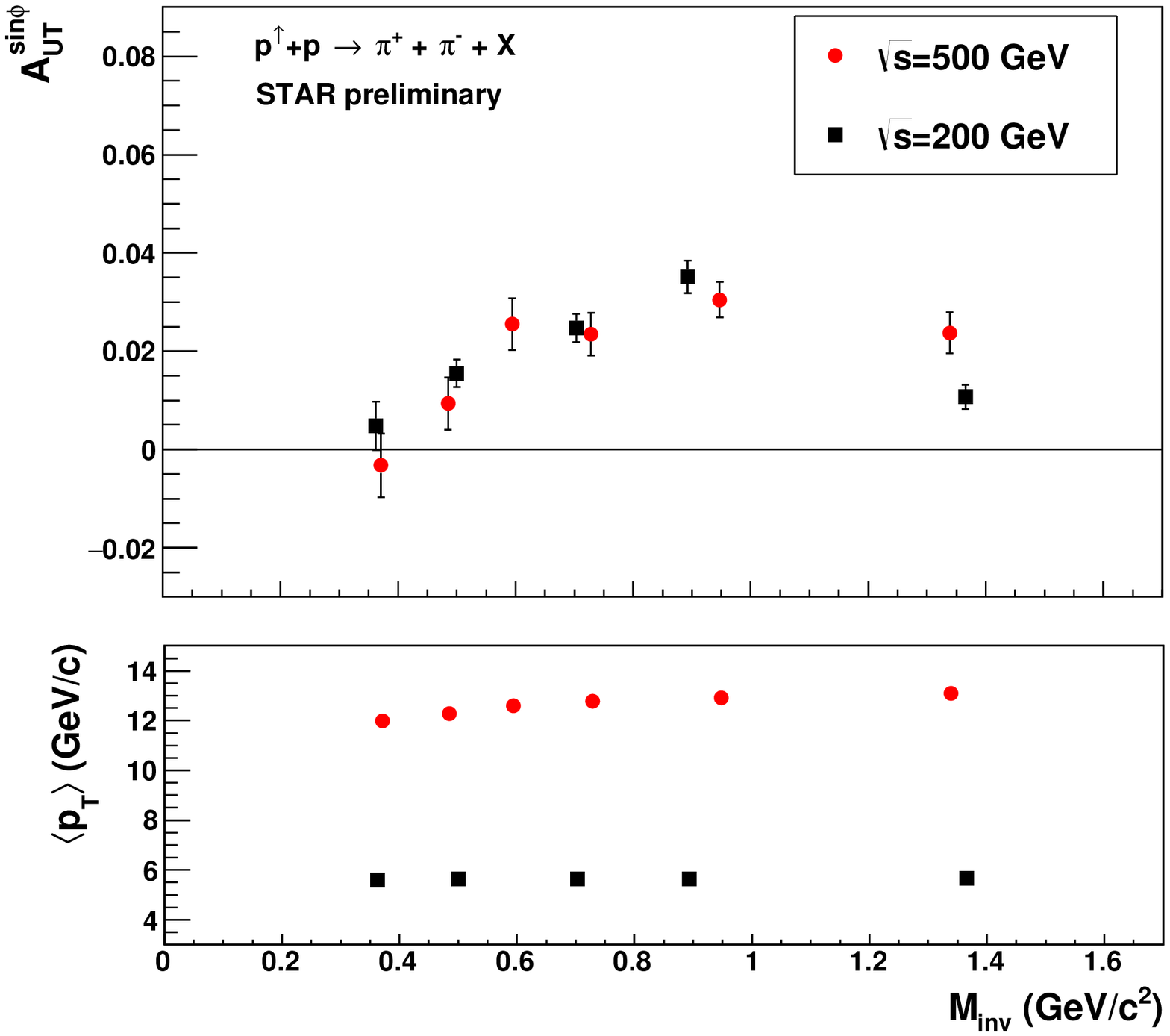} }
\caption{(a): The Collins asymmetry $A_{UT}^{\sin(\phi_S-\phi_H)}$ vs.~$z$ ($z=p_{\pi}/p_{\rm jet}$) for charged pions in jets
at $0 <\eta <1$ from $pp$ collisions at $\sqrt{s}$ = 200 and 500 GeV by STAR~\cite{Spin2014:Adkins}. The $p_{T,\rm jet}$ ranges have been chosen
to sample the same parton $x$ values for both beam energies. The angular cuts, characterised by the minimum distance of the charged pion from the jet thrust axis, have been chosen to sample the same $x_T$ values ($x_T=2p_T/\sqrt s$).
(b): The di-hadron asymmetry $A_{UT}^{\sin(\phi)}$ (upper panel) and the corresponding $p_{T{\pi^+\pi^-}}$ (lower panel) as a function of the invariant mass $M_{\pi^+\pi^-}$  for $-1 < \eta < 1$. A clear enhancement of the signal around the $\rho$-mass region
is observed both at $\sqrt{s}$ = 200 and 500 GeV by STAR~\cite{Vossen:2012zz}.}
\label{Collins.IFF}
\end{center}
\end{figure*}

\subsection{Access to Transversity: the Collins and Interference Fragmentation Functions}
\label{sec:collins}
To have a complete picture of the proton structure at leading twist one has to consider not only the unpolarised and helicity parton distributions, but also those involving transverse polarisation, as the transversity distribution. Transversity is difficult to access due to its chiral-odd nature, requiring the coupling of this distribution to another chiral-odd distribution.
Following the decomposition described in Refs.~\cite{Yuan:2007nd,Yuan:2008yv,D'Alesio:2010am,Bacchetta:2004it} the quark transversity distribution coupled to the Collins TMD or to the interference fragmentation function (IFF) may be accessed through single spin asymmetries measured in transversely polarised proton-proton collisions. Recent results from transversely polarised data taken in 2006, 2011, and 2012 at RHIC demonstrate for the first time that this is the case, thanks to the SSA measurements of the azimuthal distributions of hadrons inside a high-energy jet and of the azimuthal asymmetries of pairs of oppositely charged pions respectively at $\sqrt{s}$ = 200 and 500 GeV.

Figure~\ref{Collins.IFF} shows the first clear observations of nonzero Collins~\cite{Spin2014:Adkins} and di-hadron~\cite{Vossen:2012zz} asymmetries in $pp$ collisions at $\sqrt{s} =~200$ and 500 GeV by STAR. STAR finds that the azimuthal asymmetry of pions in polarised jet production (Fig.~\ref{Collins.IFF} (a)) depends strongly on $j_T$ $(j_T \sim \Delta R \, p_{T,\rm jet})$, the momentum of the pion transverse to the jet thrust axis.
Large asymmetries are seen for moderate values of $j_T$, whereas much smaller asymmetries are found when the measurement is restricted to larger values of $j_T$.
The measurements exhibit basically no dependence on $\sqrt{s}$ despite the jet $p_T$ differs by a factor 2, chosen to sample for both data sets the same
$x_T$ ($x_T = 2 p_T/\sqrt{s}$) to account for the $x$ dependence. These data could represent a strong constraint on the evolution of TMDs, to either show evolution effects are small or cancel to a large extent in asymmetries.
The ``IFF-asymmetries'' (Fig.~\ref{Collins.IFF}(b)) show a clear enhancement of the signal around the $\rho$-mass region. The equivalent measurement from PHENIX~\cite{Yang:2009zzr} is compatible with zero, which can be explained by the reduced rapidity ($-0.35 < \eta < 0.35$) and
transverse momentum coverage of the hadron pair and that unidentified charged hadrons have been paired with each other or with $\pi^0$.
The ``IFF-asymmetries'' measured by STAR, like the Collins asymmetries, show also no dependence on $\sqrt{s}$ for fixed $x_T$. This, in principle, is not unexpected as the ``IFF-asymmetries'' are not following TMD-evolution, but are subject to collinear evolution.
Extracting the transversity PDF from both observables, would further give the unique opportunity to test the magnitude of the theoretically predicted factorisation breaking effects for TMDs~\cite{Rogers:2013zha} measured in proton-proton collisions. For a detailed and up-to-date review on IFFs see Ref.~\cite{Pisano:2015wnq} (this Special Issue).

\begin{figure}
 \includegraphics[width=1.\columnwidth]{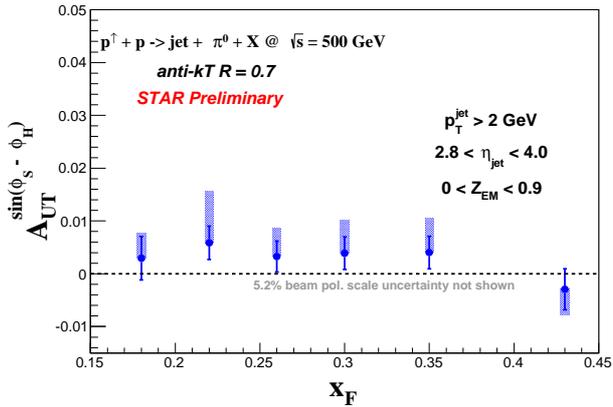}
  \caption{The Collins azimuthal asymmetry $A_{UT}^{\sin(\phi_S-\phi_H)}$ of a $\pi^0$ inside a calorimetric
   jet measured by the STAR FMS at $2.8 < \eta < 4.0$ as a function of $x_F$.}
 \label{Collins.FMS}
\end{figure}

Figure~\ref{Collins.FMS} shows a nonzero Collins asymmetry for $p^{\uparrow} p \rightarrow {\rm jet} + \pi^0 + X$ measured at $2.8 < \eta_{\rm jet} < 4.0$ \cite{Spin2014:Pan}.
The jet is reconstructed only from the calorimetric energy deposited in the STAR Forward Meson Spectrometer (FMS), an electromagnetic calorimeter at rapidity $2.8 < \eta < 4.0$. The kinematics of this measurement is identical to the one of the large forward asymmetries $A_N$ as depicted in Fig.~\ref{AN.world}, for which the Collins mechanism was one of the possible explanations, together with the Sivers effect. This measurement clearly shows that the Collins mechanism {\em alone} cannot explain the observed inclusive $\pi^0$-asymmetries, as also pointed out in Ref.~\cite{Anselmino:2012rq} (see Sec.~\ref{pph}). Fits to SIDIS and $e^+e^-$ data to extract transversity and the Collins FF find that the $u$- and $d$-quark transversity distributions are of opposite sign, with the $u$-quark transversity being larger in magnitude, and that the favoured and disfavoured Collins FFs are of similar magnitude, but of opposite sign. This, together with the fact that $\pi^0$-mesons are equally composed out of $u$- and $d$-quarks, explains the small contribution to the Collins asymmetry.

\begin{figure}
 \includegraphics[width=1.\columnwidth]{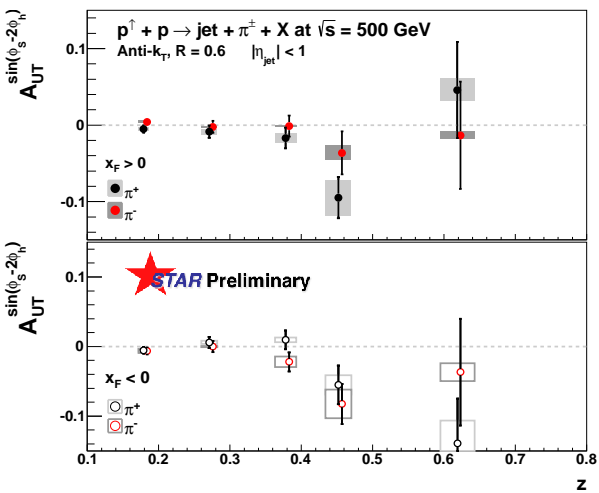}
  \caption{“Collins-like” asymmetry, $A_{UT}^{\sin(\phi_S-2\phi_h)}$, as a function of pion $z$. These results
   could provide the first experimental constraint on model predictions utilizing
   linearly polarised gluons. See also Sec.~\ref{pion-jet}.}
 \label{Collins.lG}
\end{figure}

Following the decomposition described in Ref.~\cite{D'Alesio:2010am}, that shows how to correlate different angular modulations to different TMDs, STAR has extracted several other angular modulations~\cite{Drachenberg:2014txa,Drachenberg:2014jla}. One example is the “Collins-like” asymmetry $A_{UT}^{\sin(\phi_S-2\phi_h)}$ as a function of $z$, integrated over the full acceptance and separated in forward and backward scattering relative to the polarised beam, shown in Fig.~\ref{Collins.lG}. Currently all existing model predictions are unconstrained by measurements and suggest a maximum possible upper limit of $\sim$ 2\% (see Sec.~\ref{pion-jet}). The present data fall well below this maximum with the best precision at lower values of $z$, where models suggest the largest effects may occur. Thus, these data should allow for the first phenomenological constraint on model predictions utilizing linearly polarised gluons beyond the positivity bounds.

\subsection{Transverse Polarisation in the Proton; Sivers and Qiu-Sterman effects}
\label{sec:3}
Among the quantities of particular interest to give insight into the transverse spin structure of hadrons is the “Sivers function”~\cite{Sivers:1989cc,Sivers:1990fh}, because it encapsulates the correlations between the parton transverse momentum inside the proton and the proton spin. It was found that the Sivers function, $f_{1T}^{\perp q}(x,k_{\perp}^2)$, is not universal in the hard-scattering processes, which has as its physical origin what can be described as a rescattering of the struck parton in the colour field of the remnant of the polarised proton~\cite{Brodsky:2002cx,Collins:2002kn}.
The experimental test of this non-universality is one of the big remaining questions in hadronic physics and it is deeply connected to our understanding of QCD factorisation.

RHIC provides a unique opportunity for the ultimate test of the theoretical concept of TMDs, their factorisation theorems, evolution and non-universality properties, by measuring $A_N$ for $W^{\pm}, Z^0$ bosons and Drell Yan production.
The measurement of these observables provides the two hard scales, $p_T$, the transverse momentum of the virtual boson, and $Q^2 (= M^2)$ with $p_T \ll Q$, as required in the TMD framework. The transversely polarised data set in Run-2011 at $\sqrt{s}$ = 500 GeV allowed STAR to measure the transverse single spin asymmetries, $A_N$, for $W^{\pm}$ and $Z^0$-boson production \cite{Adamczyk:2015gyk}. Especially the measurement of $A_N$ for $W^{\pm}$ bosons is challenging since, contrary to the longitudinally polarised case, it is required to completely reconstruct the $W$ bosons, as the kinematic dependences of $A_N$ cannot easily be resolved through the high-$p_T$ decaying lepton; for details see Refs.~\cite{Kang:2009bp,Metz:2010xs,Huang:2015vpy}.

\begin{figure*}[ht!]
 \begin{center}
  \subfigure[]{
   \centering
    \includegraphics[width=0.90\columnwidth]{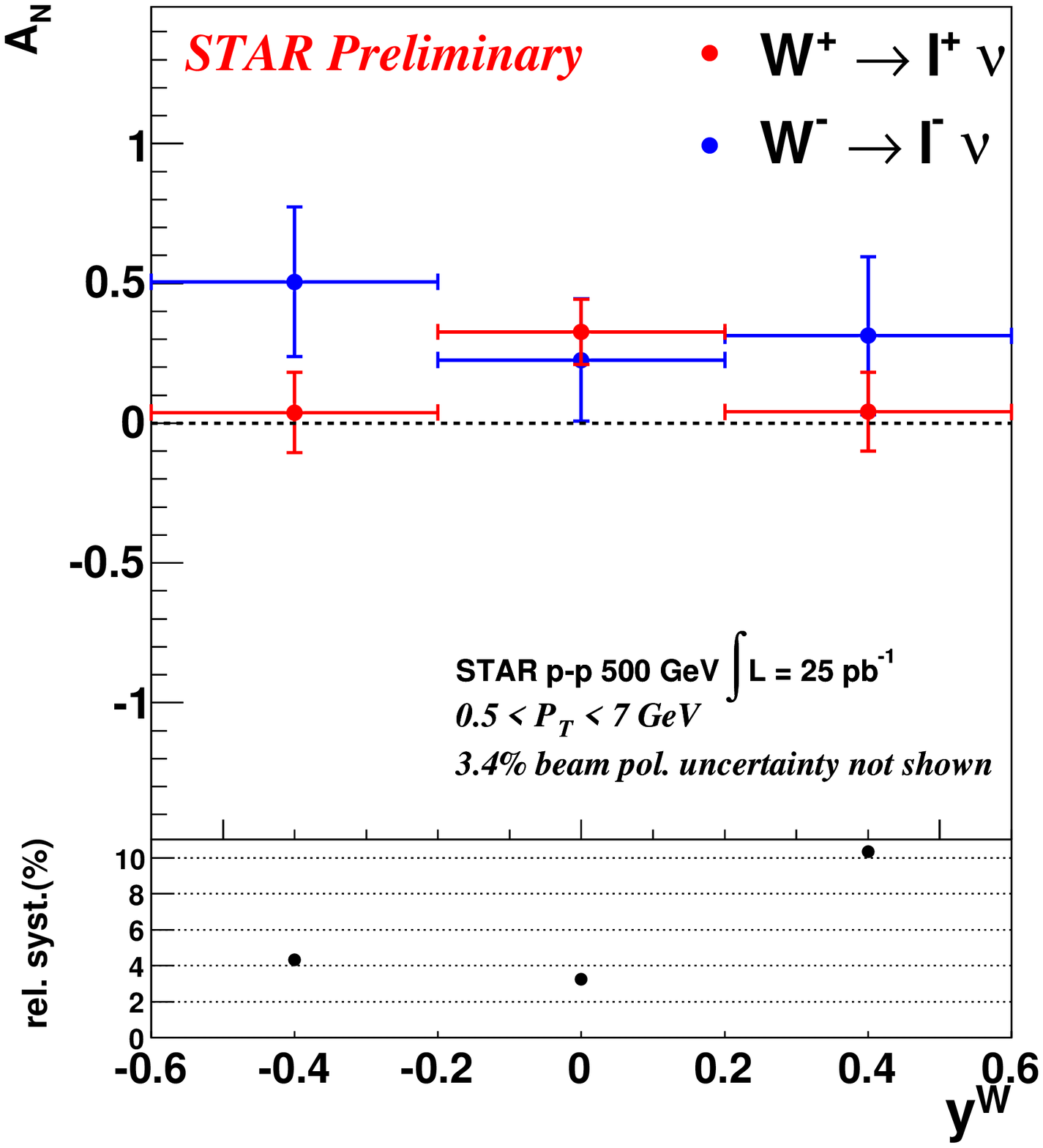} }
  \subfigure[]{
   \centering
    \includegraphics[width=0.90\columnwidth]{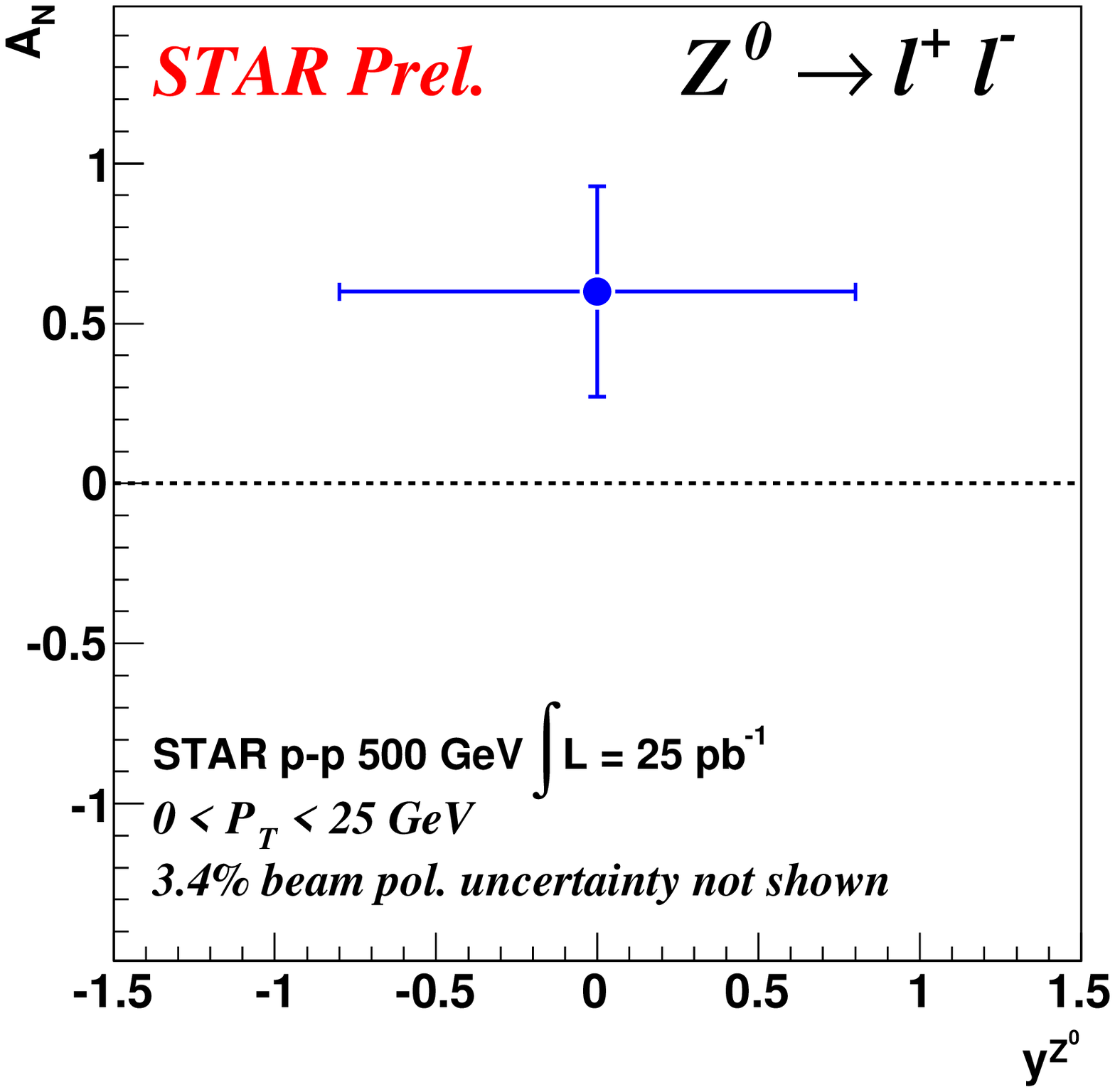} }
  \caption{The amplitude of the transverse single spin asymmetry for $W^{\pm}$ (a) and $Z^0$ (b) -boson
           production measured by STAR in proton-proton collisions at $\sqrt{s}$ = 500 GeV
           for a recorded integrated luminosity of 25 pb$^{-1}$.}
  \label{AN.W.Z0}
 \end{center}
\end{figure*}
\begin{figure*}[ht!]
 \begin{center}
   \centering
    \includegraphics[width=1.90\columnwidth]{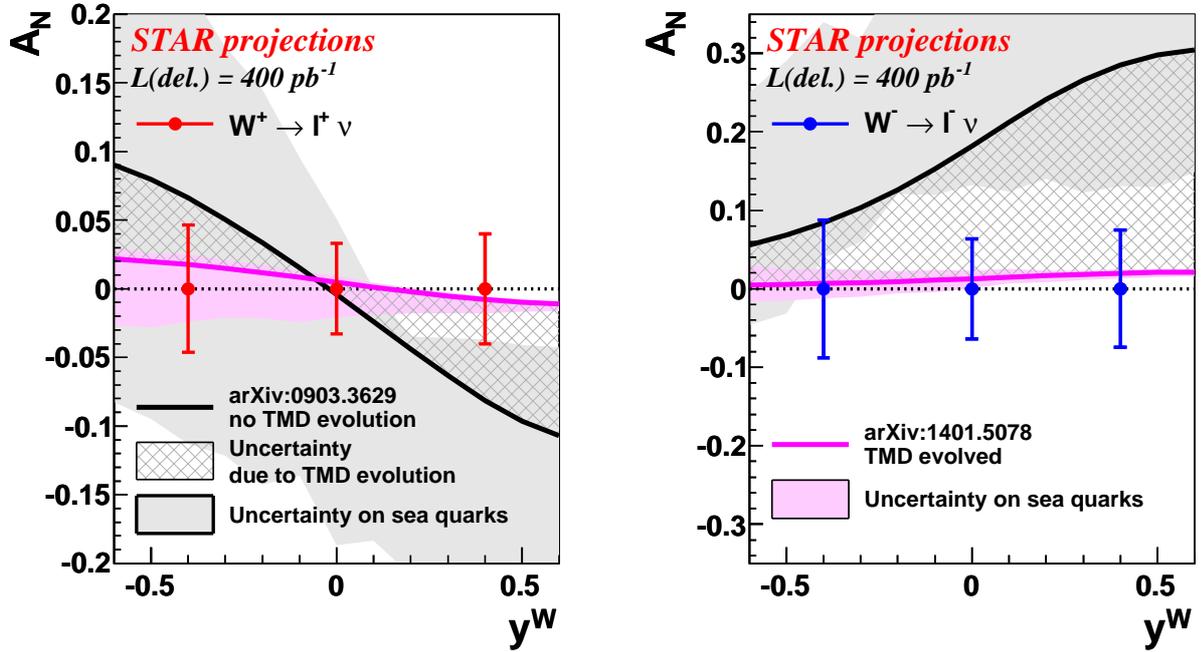}
  \caption{The projected uncertainties for transverse single spin asymmetries of $W^{\pm}$ bosons as
           a function of rapidity for a delivered integrated luminosity of 400 pb$^{-1}$ and an
           average beam polarisation of 55\%.
           The solid light gray and pink bands represent the uncertainty on the KQ~\cite{Kang:2009bp}
           and EIKV~\cite{Echevarria:2014xaa} results, respectively, due to the unknown sea-quark Sivers function.
           The crosshatched dark grey region indicates the current uncertainty in the theoretical
           predictions due to TMD evolution.}
  \label{AN.W.Z0.Proj}
 \end{center}
\end{figure*}

\begin{figure*}[ht!]
\begin{center}
\centering
  \includegraphics[width=1.80\columnwidth]{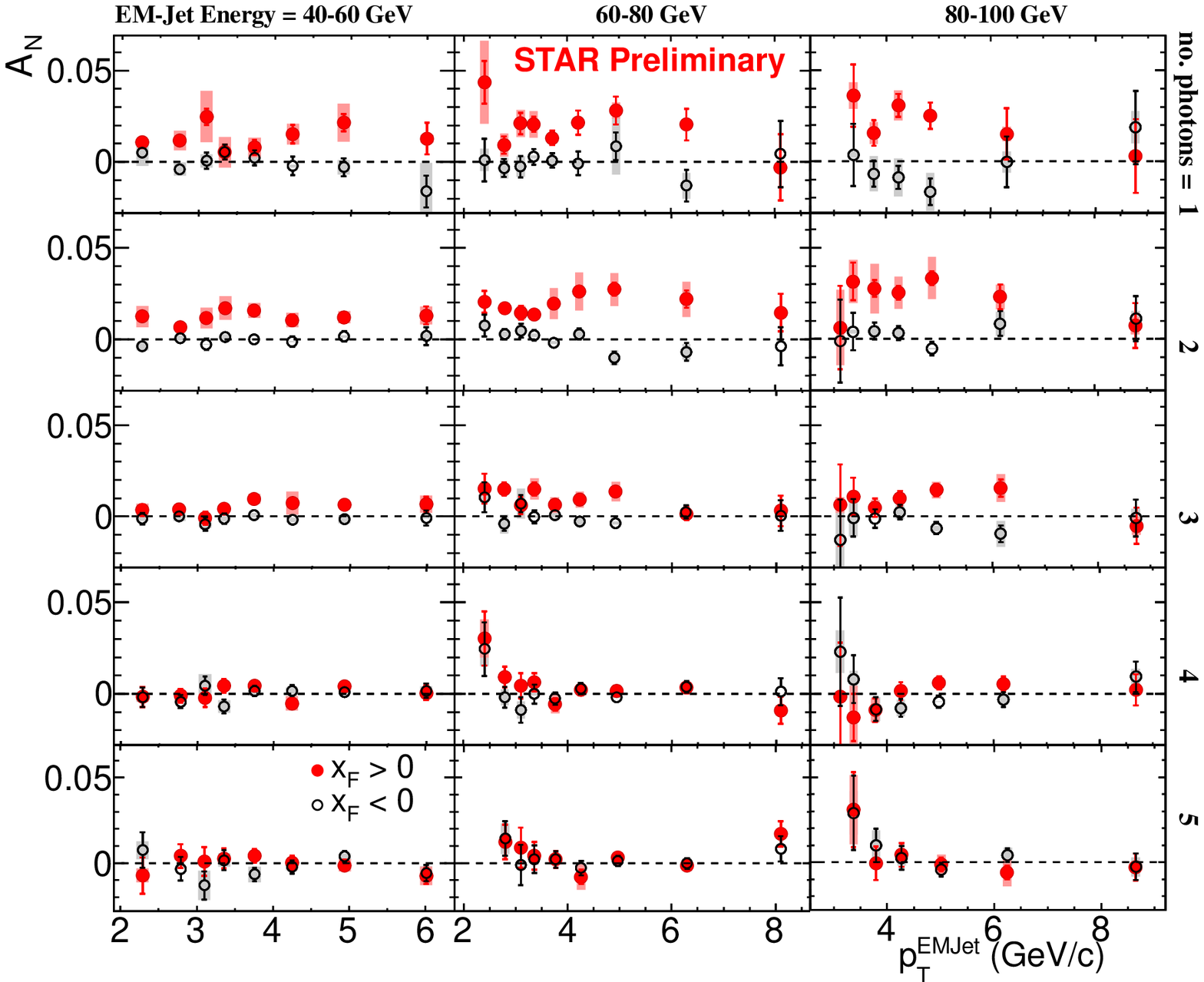}
\caption{The transverse single spin asymmetry $A_N$ for ``electromagnetic jets'' detected in the
FMS ($2.5 < \eta < 4.0$) as a function of the jet $p_T$ and the photon multiplicity in the jet
in bins of the jet energy.}
\label{FMS.pi0}
\end{center}
\end{figure*}

Thanks to the large STAR acceptance it is possible to reconstruct the $W$-boson kinematics from the recoil jet, a technique also used at D0, CDF and the LHC experiments. 
Figure~\ref{AN.W.Z0}(a) shows the transverse single spin asymmetries $A_N$ for $W^{\pm}$, as a function of the $W$-boson rapidity $y$. The asymmetries have also been reconstructed as a function of the $p_T$ of the $W$ bosons. For the $Z^0$ boson (Fig.~\ref{AN.W.Z0}(b)) the asymmetry could only be reconstructed in one $y$-bin due to the currently limited statistics (25 pb$^{-1}$).
Details for this analysis can be found in Ref.~\cite{Fazio:2014jfa,Adamczyk:2015gyk}. A combined fit to the $W^{\pm}$ asymmetries in Fig.~\ref{AN.W.Z0} based on the theoretical predictions of Kang and Qiu (KQ)~\cite{Kang:2009bp} favors a sign-change for the Sivers function relative to the Sivers function in SIDIS by more than $2 \sigma$, if TMD evolution effects are small (see also Sec.~\ref{DY}).
This analysis represents an important proof of principle. Figure~\ref{AN.W.Z0.Proj} shows the projected uncertainties for transverse single spin asymmetries for $W^{\pm}$ bosons as a function of rapidity for a delivered integrated luminosity of 400 pb$^{-1}$ and an average beam polarisation of 55\%.
Such a measurement will provide the world wide first constraint on the light sea-quark Sivers function. At the same time, it will also give access to the sign change of the Sivers function, if the effect due to TMD evolution on the asymmetries is in
the order of a factor 5 reduction. The expected statistical precision for the $Z^0$ and DY $A_N$ are 3 points in rapidity with an uncertainty of 0.2 and one point with an uncertainty of 0.008, respectively.

At this point it should be mentioned that, in principle, most of the other observables, namely those for single inclusive particle production in $pp$ collisions, where only one hard scale is present, can be more naturally related to the transverse spin structure of hadrons through the twist-3 formalism. This scale is typically the $p_T$ of a produced particle or jet, which at RHIC is sufficiently large in much of the phase space. On the other hand the GPM approach applied to this class of observables still represents a very successful phenomenological approach and it is worth to be further explored.

For the sake of completeness we recall here that the Sivers function is related to its twist-3 counterpart, the Efremov-Teryaev-Qiu-Sterman (ETQS) function $T_{q,F}$~\cite{Efremov:1981sh,Efremov:1984ip,Qiu:1991pp,Qiu:1991wg,Qiu:1998ia}, through the following relation~\cite{Boer:2003cm,Kang:2011hk},
\begin{equation}
T_{q,F}(x,x)=-\int d^2k_{\perp} \frac{|k_{\perp}|^2}{M}f_{1T}^{\perp q}(x,k_{\perp}^2)|_{\rm SIDIS}\,.
\label{eq.twist3.sivers}
\end{equation}

\begin{figure*}
 \begin{center}
  \subfigure[]{
   \centering
    \includegraphics[width=0.90\columnwidth]{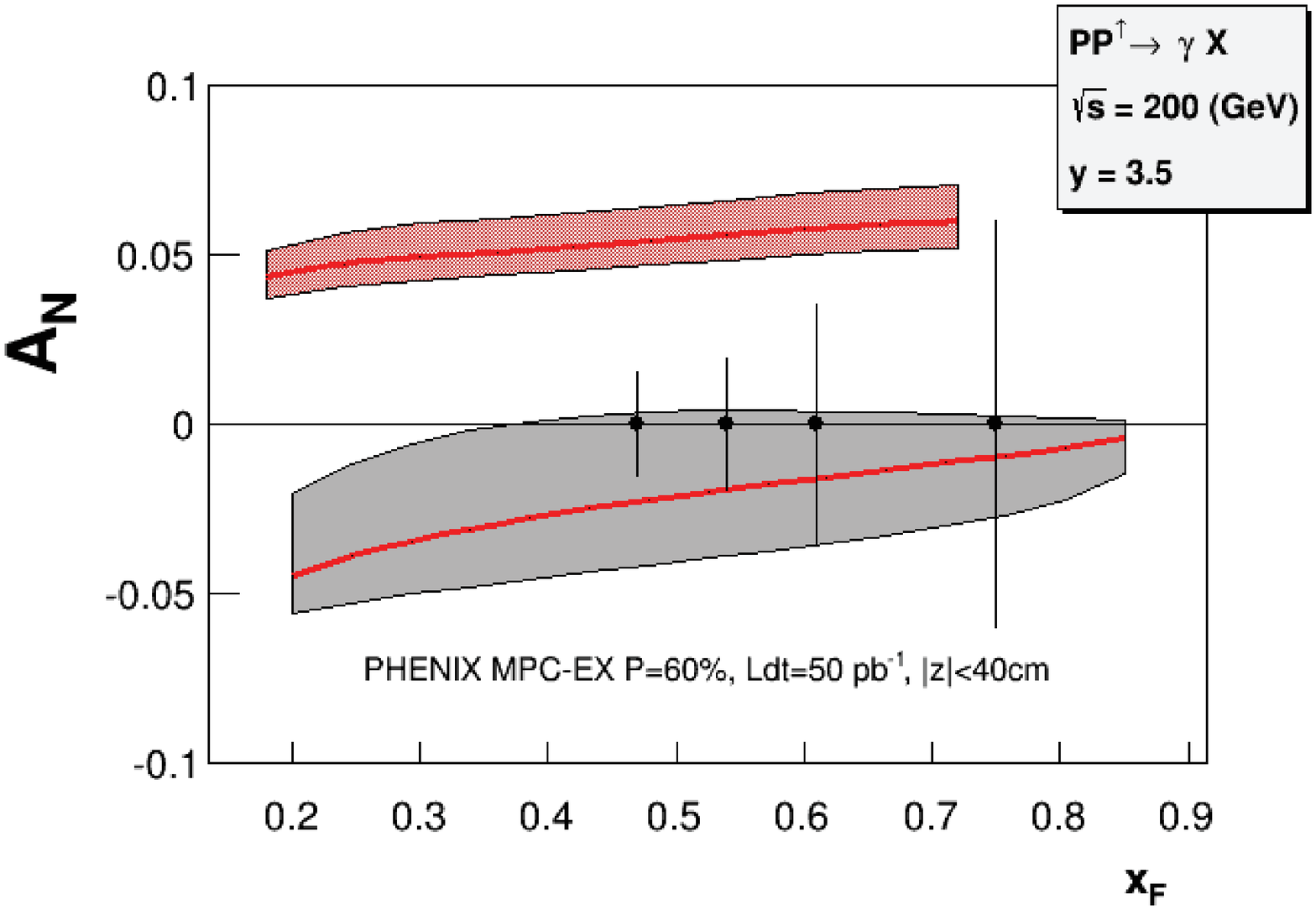} }
  \subfigure[]{
   \centering
    \includegraphics[width=0.90\columnwidth]{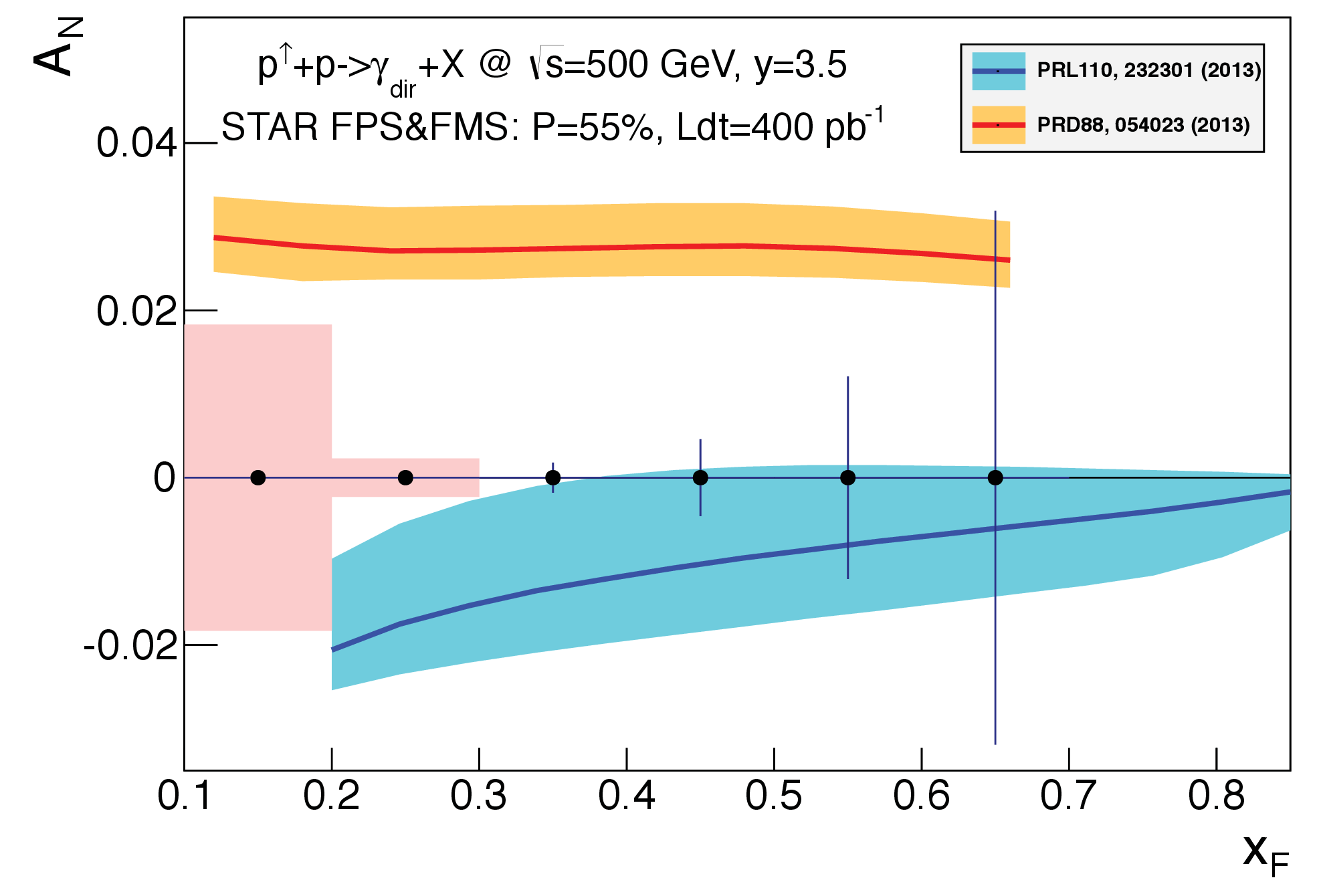} }
  \caption{Statistical and systematic uncertainties for the direct photon $A_N$ after background subtraction compared to theoretical predictions from Ref.~\cite{Anselmino:2013rya} in the GPM and Ref.~\cite{Gamberg:2013kla} in the twist-3 approach for $\sqrt{s}$ = 200 GeV as measured by PHENIX (a) and for $\sqrt{s}$ = 500 GeV as measured by STAR (b).}
  \label{direct.g.P.S.}
 \end{center}
\end{figure*}

The primary observable of PHENIX and STAR to transverse spin physics has been through the study of forward neutral pion production in $pp$ collisions (see, for example, Refs.~\cite{Adamczyk:2013yvv,Abelev:2008af}). This effort has been extended to include the first measurements at $\sqrt{s}$ = 200 GeV of the transverse spin asymmetry $A_N$ for the $\eta$ meson~\cite{Adamczyk:2012xd}.
The STAR Run-2011 data taken with transverse polarisation at $\sqrt{s}$ = 500 GeV have revealed several surprising results.
Figure~\ref{FMS.pi0} shows the transverse single spin asymmetry $A_N$ for ``electromagnetic jets'' (i.e. jets with their energy only measured in an electromagnetic calorimeter) detected in the FMS at $2.5 < \eta < 4.0$ as a function of the jet $p_T$ for different photon multiplicities and ranges in jet energy~\cite{Mondal:2014vla}.
It can be clearly seen that with increasing number of photons in the ``electromagnetic jet'' the asymmetry decreases. Jets with an isolated $\pi^0$ exhibit the largest asymmetry, consistent with the asymmetry in inclusive $\pi^0$ events, as seen from the right-most panel in Fig.~\ref{AN.world}. For all jet energies and photon multiplicities in the jet, the asymmetries are basically flat as a function of jet $p_T$, a feature also already seen for the inclusive $\pi^0$ asymmetries. This behaviour is very different from what would be naively expected for an asymmetry driven by QCD subprocesses, which would follow a 1/$p_T$ dependence (see discussion in Sec.~\ref{pph}).

This STAR result is in agreement with preliminary observations from the A$_N$DY collaboration at RHIC, which measured $A_N$ for inclusive jets at $\sqrt{s}$ = 500 GeV in the order of $\sim 5 \times 10^{-3}$~\cite{Bland:2013pkt}. All these observations might indicate that the underlying subprocess causing a significant fraction of the large transverse single spin asymmetries in the forward direction are not of $2 \rightarrow 2$ parton scattering processes but of diffractive nature.
During the 2015 transversely polarised $pp$ run at $\sqrt{s}$ = 200 GeV this conjecture can be definitely tested by STAR measuring $A_N(\pi^0)$ for single and double diffractive events by tagging one or both protons in the Roman Pots.

Transverse single spin asymmetries in direct photon production provide a very interesting and powerful tool: they could indeed represent a clear observable allowing to disentangle the different mechanisms at work in the initial state (see Sec.~\ref{photon-jet}). For the 2015 polarised $pp$ run both PHENIX and STAR have installed preshowers in front of their forward electromagnetic calorimeters, the  muon piston calorimeter (MPC) and the FMS, respectively~\cite{PHENIX:BUR2015,STAR:BUR2015}.
These upgrades will enable a measurement of the SSA for direct photons both at $\sqrt{s}$ = 200 GeV and 500 GeV in the same kinematics where the inclusive $A_N(\pi^0)$ is largest, i.e.~up to $x_F \sim 0.7$.
For these asymmetries no potential cancelation as for jets and $\pi^0$ due to the opposite sign but similar magnitude of the $u$ and $d$ quark contributions is expected, since the electromagnetic nature of the process implies that the individual parton densities are weighted with the respective quark charge $e_q^2$. Figure~\ref{direct.g.P.S.}(a) shows the statistical and systematic uncertainties for the direct photon $A_N$ as to be obtained by PHENIX in Run-2015 at $\sqrt{s}$ = 200 GeV; similar uncertainties will be obtained by STAR. The asymmetry can be measured up to $x_F \sim$ 0.7 where the inclusive $\pi^0$ asymmetries are largest (see Fig.~\ref{AN.world}). The curves represent calculations based on the GPM approach~\cite{Anselmino:2013rya} (see also Sec.~\ref{photon-jet} for further details) and the twist-3 formalism~\cite{Gamberg:2013kla}.  Figure~\ref{direct.g.P.S.}(b) shows the same simulation and uncertainties for the direct photon SSA at $\sqrt{s}$ = 500 GeV, a measurement earliest possible in 2017. The theoretical asymmetries are reduced by a factor 2 due to the larger $p_T$ values spanned at $\sqrt{s}$ = 500 GeV w.r.t.~the corresponding ones at 200 GeV for the GPM and to the collinear evolution in the twist-3 approach. The precision of the data will be good enough to separate the two predictions at both center-of-mass energies.

\begin{figure}[ht!]
  \includegraphics[width=0.90\columnwidth]{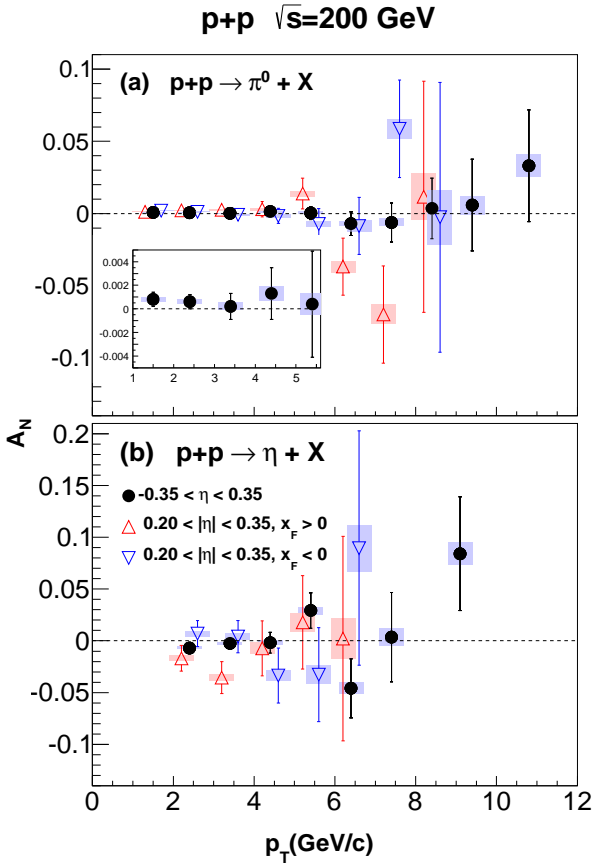}
\caption{$A_N$ measured at midrapidity ($|\eta| < 0.35$), as a function of $p_T$ for $\pi^0$ (upper plot) and $\eta$ (lower plot) mesons. Triangles are slightly forward/backward going sub-samples of the full data set (circles). These are shifted in $p_T$ for better visibility. An additional
uncertainty from the beam polarisation is not included.}
\label{Twist3.gluon.pi0}
\end{figure}

Both PHENIX and STAR have made important measurements at midrapidity suitable to constrain, in particular, gluon TMDs. Figure~\ref{Twist3.gluon.pi0} shows $A_N$ for $\pi^0$ (a) and $\eta$ (b) mesons at midrapidity at $\sqrt{s}$ = 200 GeV. No significant deviation from zero can be seen in the results within the statistical uncertainties in the covered transverse momentum range. With the dominant underlying subprocesses being $qg$ and $gg$ scattering, these asymmetries can be used to constrain the gluon Sivers function within the GPM approach~\cite{Anselmino:2006yq,D'Alesio:2015uta} (see Sec.~\ref{pph}).
PHENIX has further made measurements of $A_N$ for inclusive muons~\cite{PHENIX.MUON} and $J/\psi$~\cite{Adare:2010bd} mesons measured at $1.4 <|\eta|<1.9$ with the PHENIX muon arms. These inclusive muons come dominantly from open charm and  $J/\psi$ mesons, which can be only produced through gluon-gluon fusion and therefore give a direct sensitivity to the integral over $x$ of the gluon Sivers function within the GPM and the three-gluon correlator within the twist-3 formalism.
The transversely polarised STAR data taken at $\sqrt{s}=500$ GeV in 2011 also provide enhanced sensitivity to lower $x$ partons and therefore distributions such as the gluon Sivers TMD.

\begin{figure}
  \includegraphics[width=0.98\columnwidth]{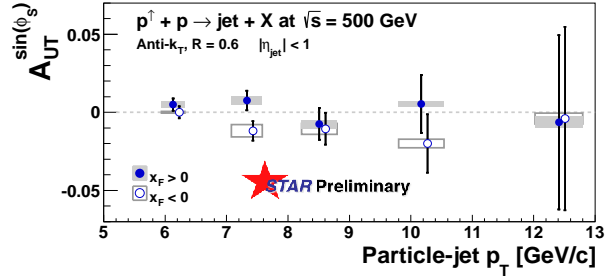}
\caption{Inclusive jet azimuthal transverse single spin asymmetries as a function of
particle jet $p_T$ for $-1 < \eta_{\rm jet} < 1$ relative to the polarised beam. The measurement shows
no sign of large asymmetries and may suggest further constraints on the gluon Sivers and/or the corresponding twist-3 three-gluon correlator.}
\label{Twist3.gluon.jet}
\end{figure}

The Sivers function may be accessed via the azimuthal asymmetries of inclusive jets. These asymmetries, shown in Fig.~\ref{Twist3.gluon.jet}, are presented as a function of particle-jet $p_T$ for $-1 < \eta_{\rm jet} < 1$. No large asymmetries are observed, consistently with expectation from measurements at $\sqrt{s}$ = 200 GeV~\cite{Abelev:2007ii,Adamczyk:2012qj,Adare:2013ekj} and model predictions~\cite{Kanazawa:2012kt,D'Alesio:2015uta}.

\section{Theory and Phenomenology}

In this Section we will present a selection of results within the GPM approach on SSAs for single-inclusive large-$p_T$ pion, jet and photon and for double-inclusive pion-jet production in proton-proton collisions, as well as for Drell-Yan processes. A brief account of more general double inclusive processes is also given. For a complete treatment and discussion of all these cases we refer to the related quoted papers. 

\subsection{SSAs in $p^\uparrow p\to h\, X$ processes}
\label{pph}
This process still represents a puzzling issue in QCD. In the GPM $A_N$ originates mainly from two spin and transverse momentum dependent effects, one introduced by Sivers in the partonic distributions~\cite{Sivers:1989cc,Sivers:1990fh}, and one by Collins in the parton fragmentation
process~\cite{Collins:1992kk}. All other TMD effects are in fact strongly suppressed by intrinsic azimuthal phase integrations~\cite{Anselmino:2005sh}.

As already mentioned the transverse single spin asymmetry $A_N$, measured in $\pup p\to h \, X$ inclusive reactions is defined as:
\be
A_N = \frac{d\sigma^\uparrow - d\sigma^\downarrow}
           {d\sigma^\uparrow + d\sigma^\downarrow}
\quad\, {\rm with} \quad\,
d\sigma^{\uparrow, \downarrow} \equiv
\frac{E_h \, d\sigma^{p^{\uparrow, \downarrow} p\to h \, X}}
{d^{3} \bfp_h} \>, \label{an}
\ee
where $\uparrow, \downarrow$ stay for opposite spin orientations perpendicular to the $x$-$z$ scattering plane, in the $\pup p$ c.m.~frame. We define the $\uparrow$ direction as the $+\hat y$-axis, with the polarised proton moving along the $+\hat z$-direction. Notice that $d\sigma^\uparrow+ d\sigma^\downarrow$ is twice the unpolarised cross section.
In such a process the relevant large energy scale is the transverse momentum $p_T=|(\bfp_h)_x|$ of the final hadron.

According to the GPM formalism~\cite{Anselmino:2005sh,Anselmino:2012rq,Anselmino:2013rya}, $A_N$ can then be written as:
\be
A_N \cong \frac{[d\sigma^\uparrow - d\sigma^\downarrow]_{\rm Sivers}
+ [d\sigma^\uparrow - d\sigma^\downarrow]_{\rm Collins}}
{d\sigma^\uparrow + d\sigma^\downarrow} \>\cdot \label{ansc}
\ee
The Collins and Sivers contributions, which were recently reconsidered respectively in Refs.~\cite{Anselmino:2012rq} and \cite{Anselmino:2013rya}, are given in the GPM by:
\begin{align}
[d\sigma^\uparrow & - d\sigma^\downarrow]_{\rm Sivers}
=   \sum_{abcd} \int d[PS] \nonumber\\
& \times \Delta^N\!f_{a/\pup}(x_a, k_{\perp a}) \,
\cos (\phi_a) \,f_{b/p}(x_b, k_{\perp b}\nonumber )\\
& \times   \frac{1}{2}
\left[ |\hat M_1^0|^2 + |\hat M_2^0|^2 + |\hat M_3^0|^2 \right]
D_{h/c}(z, p_\perp) \label{numans}
\end{align}
and
\begin{align}
[d\sigma^\uparrow & - d\sigma^\downarrow]_{\rm Collins}
= \sum_{q_a,b,q_c,d} \int d[PS] \nonumber \\
& \times  \Delta_Tq_a(x_a, k_{\perp a}) \,
\cos (\phi_a + \varphi_1 - \varphi_ 2 + \phi_h^H) \nonumber \\
& \times f_{b/p}(x_b, k_{\perp b}) \>
\left[ \hat M_1^0 \, \hat M_2^0 \right] \>
\Delta^N D_{h/\qup_c}(z, p_\perp) \>, \label{numanc}
\end{align}
where $d[PS]$ is the proper phase-space factor including all relevant kinematical factors and the $\hat M_i^0$'s are the partonic c.m.~helicity amplitudes.
For details and a full explanation of the notation in the above equations see Ref.~\cite{Anselmino:2005sh} (where $\bfp_\perp$ is denoted as
$\bfk_{\perp C}$).

The Sivers effect is encoded in the number density of unpolarised quarks $q$ (or gluons) with
light-cone momentum fraction $x$ and intrinsic transverse momentum $\bfk_\perp$ inside a transversely polarised proton $\pup$, with three-momentum $\bfP$ and spin polarisation vector $\bfS$, that can be written as
\be
\hat f_ {q/\pup} (x,\bfk_\perp) = f_ {q/p} (x,\kt) +
\frac{1}{2} \, \Delta^N \! f_ {q/\pup}(x,\kt) \;
{\bfS} \cdot (\hat {\bfP}  \times \hat{\bfk}_\perp)
\,,\label{sivnoi}
\ee
where $f_ {q/p}(x,\kt)$ is the unpolarised TMD PDF ($\kt = |\bfk_\perp|$), $\Delta^N \! f_ {q/\pup}(x,\kt)$ is the Sivers function, and $\hat {\bfP} = \bfP/|\bfP|$, $\hat{\bfk}_\perp = \bfk_\perp/\kt$ are unit vectors.
Notice that the Sivers function is also denoted as $f_{1T}^{\perp q}(x, k_\perp)$~\cite{Mulders:1995dh,Bacchetta:2004jz}.

Similarly, due to the Collins effect, the number density of unpolarised hadrons $h$ with light-cone momentum fraction $z$ and transverse momentum $\bfp_\perp$ produced in the fragmentation of a transversely polarised quark $\qup$, with three-momentum $\bfq$ and spin polarisation vector $\bfS_q$, can be written as
\be
\hat D_{\qup/h} (z,\bfp_\perp) = D_{h/q} (z,\pt) +
\frac{1}{2} \, \Delta^N \! D_{\qup/h}(z,\pt) \,
{\bfS_q} \cdot (\hat {\bfq}  \times \hat{\bfp}_\perp) ,
\label{colnoi}
\ee
where $D_ {h/q}(z,\pt)$ is the unpolarised TMD FF ($\pt = |\bfp_\perp|$), $\Delta^N \! D_ {\qup/h}(z,\pt)$ is the Collins function, and $\hat {\bfq} = \bfq/|\bfq|$,  $\hat{\bfp}_\perp = \bfp_\perp/\pt$ are unit vectors.
Notice that the Collins function is also denoted as $H_{1}^{\perp q}(z, p_\perp)$~\cite{Mulders:1995dh,Bacchetta:2004jz}.

The phase factor $\cos(\phi_a)$ in Eq.~(\ref{numans}) originates directly from the $\bfk_\perp$ dependence of the Sivers distribution [${\bfS} \cdot (\hat {\bfP} \times \hat{\bfk}_\perp)$, Eq.~(\ref{sivnoi})]. Analogously, the phase factor $\cos(\phi_a + \varphi_1 - \varphi_ 2 + \phi_h^H)$ in Eq.~(\ref{numanc}) originates from the $\bfk_\perp$ dependence of the unintegrated transversity distribution $\Delta_Tq$, the polarised elementary interaction and the spin-$\bfp_\perp$ correlation in the Collins function. The explicit expressions of $\varphi_1$, $\varphi_2$ and $\phi_h^H$ in terms of the integration variables can be found via Eqs.~(60)-(63) in~\cite{Anselmino:2005sh} and Eqs.~(35)-(42) in~\cite{Anselmino:2004ky}.

For the process $p^\uparrow p\to \pi\, X$ the Collins and Sivers effects cannot be disentangled: both of them could, in principle, contribute to $A_N$. While a global fit including all available data is still premature, one could start showing to what extent the corresponding TMDs, so far extracted from fits to azimuthal asymmetries measured in SIDIS and $e^+e^-$ annihilation processes, are able to describe $A_N$ data. To this aim we consider separately the two effects as acting alone. In the following we will focus on single pion production.

Let us start by considering the forward rapidity region, where one can safely neglect the gluon Sivers contribution.

In such a case one has to take into account that SIDIS data, from which the quark Sivers function and the transversity distribution (entering the Collins effect) can be extracted, are limited to $x_B$ values lower than 0.3. This implies that, being $x>x_F$ (with $x$ the light-cone momentum fraction of the parton inside the polarised proton), the use at large $x_F$ of parameterisations coming from SIDIS fits is not completely under control.
For this reason a different strategy has been devised: one first starts with a large set of parameterisations scanning the large-$x$ behaviour of these two functions, keeping only those sets that guarantee a statistically sound fit of SIDIS data;
then, among the left sets, we select the one that describes better also the large-$x_F$ $A_N(p^\uparrow p\to \pi\,X)$ data (namely from STAR and BRAHMS experiments at $\sqrt{s}=200$ GeV).
To better substantiate the significance of these results we furthermore calculate their statistical uncertainty bands following the procedure described in Appendix A of Ref.~\cite{Anselmino:2008sga} (for a more comprehensive discussion of the entire procedure see Refs.~\cite{Anselmino:2012rq} and~\cite{Anselmino:2013rya}).
\begin{figure}[ht!]
\resizebox{0.55\textwidth}{!}{%
  \includegraphics{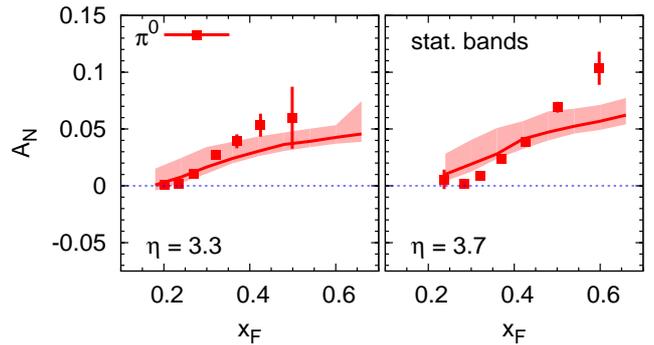}
}
\caption{The Sivers contribution to the neutral pion single spin asymmetry $A_N$, as a function of $x_F$, compared with STAR data at two fixed pion rapidities and $\sqrt s = 200$ GeV~\cite{Abelev:2008af}. The central lines are the best found curves according to the procedure delineated in the text. The shaded statistical error bands are generated applying the procedure described in Appendix A of Ref.~\cite{Anselmino:2008sga}.}
\label{fig:an-pion-Sivers}
\end{figure}
\begin{figure}[ht!]
\resizebox{0.55\textwidth}{!}{%
  \includegraphics{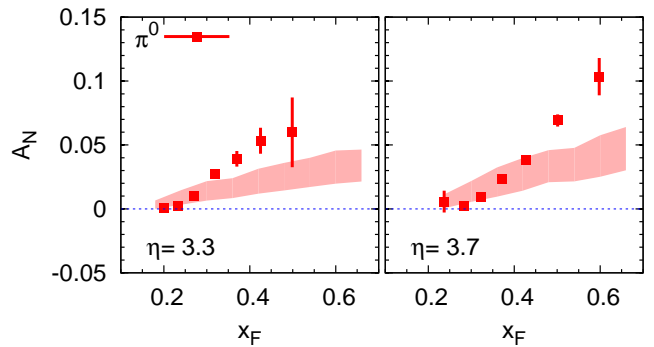}
}
\caption{The Collins contribution to the neutral pion single spin asymmetry $A_N$, as a function of $x_F$, compared with STAR data at two fixed pion rapidities and $\sqrt s = 200$ GeV~\cite{Abelev:2008af}. The shaded bands correspond to the statistical uncertainties generated following Appendix A of Ref.~\cite{Anselmino:2008sga}.
}
\label{fig:an-pion-Collins}
\end{figure}

In Figs.~\ref{fig:an-pion-Sivers} and \ref{fig:an-pion-Collins} we show our estimates for the Sivers and the Collins contributions to $A_N$ for neutral pion production, as a function of $x_F$, at $\sqrt{s}=200$ GeV and two fixed rapidity values, $\eta=3.3$ and 3.7, compared with STAR data~\cite{Abelev:2008af}. These results are obtained adopting the Kretzer set for the collinear unpolarised fragmentation functions~\cite{Kretzer:2000yf}; similar results can be obtained with the set by de Florian, Sassot and Stratmann (DSS)~\cite{deFlorian:2007aj}.
Notice that for the Collins effect we do not show the best curve since its contribution is always quite below the large $x_F$ data. From this study one can conclude that while the Sivers effect alone, as extracted from SIDIS data, could be able to describe fairly well pion $A_N$ data, the Collins effect underestimates systematically the large $x_F$ data. It remains to check whether a possible combination of the two effects could eventually describe these SSAs and, at the same time, other SSAs in $pp$ collisions where only the Sivers effect or the Collins one (see the next Sections) play a role.

Concerning the twist-3 approach, it has been shown that, by using the relations among moments of TMDs and the twist-3 functions, see  \textit{e.g.}~Eq.~(\ref{eq.twist3.sivers}), a global description of SIDIS and SSAs data is possible~\cite{Kanazawa:2014dca}. In particular, in this approach a significant portion of the sizeable inclusive pion asymmetries seen at forward pseudorapidity is due to an extra twist-3 piece in the fragmentation (that has no counterpart in the TMD sector) coupled to transversity~\cite{Kanazawa:2014dca}. This calculation shows, similar to the experimental results (see comments in Sec.~\ref{sec:3}), a flat $p_T$ dependence for $A_N$. We notice here that this flat and puzzling behaviour in $p_T$ appears also in the GPM, as discussed in Ref.~\cite{Anselmino:2013rya}.
Taking into account that both approaches are based on a perturbative QCD scheme, this result could be ascribed to a partial mitigation of the proper $1/p_T$ behaviour (intrinsic by definition in the twist-3 approach) via a compensation among different factors. Even though the twist-3 findings certainly represent a promising attempt to describe adequately the effects seen in SIDIS and in $pp$ collisions, some aspects still need to be better understood, like the role played in other processes by the new (and relevant) twist-3 fragmentation contribution. As a matter of fact currently the forward rapidities ($\eta >  2.5$) are the only kinematic region where the effects due to this twist-3 FF coupled to transversity are sizeable.

\begin{figure}
\resizebox{0.5\textwidth}{!}{%
  \includegraphics{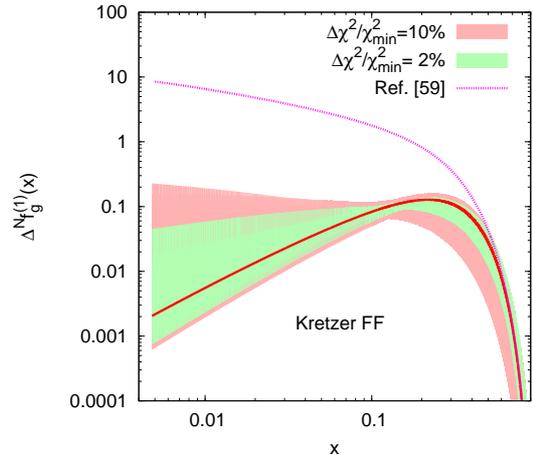}
}
\caption{First $\bm{k}_\perp$-moment of the gluon Sivers function at $Q^2=2$ GeV$^2$. The best estimate (red solid line) is shown together with the tolerance bands corresponding to a 2\% (narrower, green) and 10\% (wider, red) variation in the $\chi^2$. The former bound on the gluon Sivers function (magenta dotted line), obtained in Ref.~\cite{Anselmino:2006yq}, is also shown.}
\label{fig:gsf}
\end{figure}

Moving now to the central rapidity region, it can be shown that only the Sivers contribution could play a role. In this case, in fact, the Collins effect is suppressed by the integration over the complex azimuthal phase appearing in Eq.~(\ref{numanc}).
By using information at our disposal on the quark Sivers distribution, together with available data on $A_N(\pi^0)$ by the PHENIX Collaboration (Fig.~\ref{Twist3.gluon.pi0}, upper plot) -- almost compatible with zero -- one can gain some knowledge on the poorly known gluon Sivers function (GSF). Recently, in Ref.~\cite{D'Alesio:2015uta} a first preliminary extraction of the GSF has been presented, significantly improving the former bound~\cite{Anselmino:2006yq}.
In Fig.~\ref{fig:gsf} we present these findings, showing the best-fit result of the first $\bm{k}_\perp$ moment of the GSF (solid line) as a function of $x$ at $Q^2=2$ GeV$^2$; the dashed bands represent the possible uncertainties on this extraction by allowing for a variation of the $\chi^2$ of 2\% (green band) and 10\% (red band) w.r.t.~the minimum best fit value. Notice that this scenario is obtained with the Kretzer FF set~\cite{Kretzer:2000yf}; for a more complete discussion on this extraction see Ref.~\cite{D'Alesio:2015uta}.

Still focusing on SSAs in hadronic processes, we have to mention that to access the GSF other channels have been also proposed, like the inclusive photon production in the large negative $x_F$ region~\cite{Schmidt:2005gv}, the back-to-back azimuthal correlations in two-jet production~\cite{Boer:2003tx} and the inclusive $D$ meson production at RHIC~\cite{Anselmino:2004nk}. See Ref.~\cite{Boer:2015vga} for a detailed and updated discussion on the gluon Sivers function. Similarly, the role of linearly pola\-ri\-sed gluons inside (un)polarised protons in inclusive processes has been actively investigated, \textit{e.g.}~in pion-jet production~\cite{D'Alesio:2010am} (see Sec.~\ref{pion-jet}), heavy quark and jet-pair production at electron-ion or hadron colliders~\cite{Boer:2010zf,Pisano:2013cya}, and Higgs production at the LHC~\cite{Boer:2011kf,Boer:2014tka,Echevarria:2015uaa}.

\subsection{SSAs in $pp\to {\rm jet}\, X$ and in $pp \to \gamma \, X$}
\label{photon-jet}

In these processes no fragmentation mechanism is present, so that, within the GPM approach, one can access directly the Sivers effect, as discussed, \textit{e.g.}, in Refs.~\cite{D'Alesio:2004up,D'Alesio:2010am}. Notice that, for the same reason, SSAs for inclusive jet or photon production can be used to test the process dependence of the Sivers functions  and/or to discriminate between the GPM and the twist-3 approach.

The numerator of $A_N$ for inclusive jet production can be obtained from Eq.~(\ref{numans}) by simply replacing the TMD fragmentation function, $D_{h/c}(z,p_\perp)$, with a factor $\delta(z-1)\,\delta^2(\bm{p}_\perp)$ (and identifying now the final hadron momentum, $\bfp_h$, with the jet momentum $\bfp_{\rm jet} \equiv \bfp_c$). Notice that the elementary hard scattering interactions are exactly the same as those for inclusive hadron production and, at leading order, the jet is identified with the final parton~$c$.

Concerning direct photon production, the basic partonic processes are the Compton process $g\,q\,(\bar q) \to \gamma \, q\,(\bar q)$ and the annihilation process $q \, \bar q\to \gamma \, g$. Also in this case one can formally use the same equation (as for the inclusive jet production), by replacing the partonic helicity amplitudes with the corresponding ones for the process $a \, b \to \gamma \, d$ (see also Ref.~\cite{D'Alesio:2004up}).

\begin{figure}[ht!]
\resizebox{.45\textwidth}{!}{%
\centering
  \includegraphics{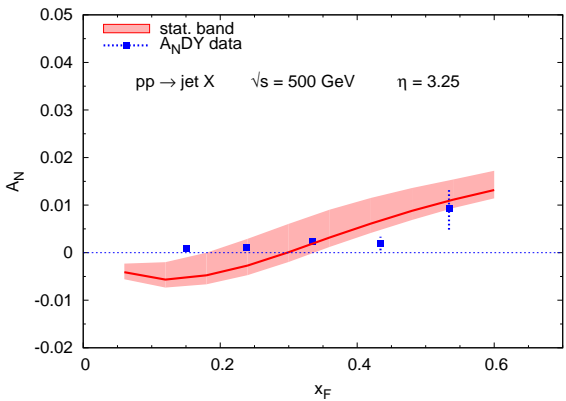}
}
\caption{$A_N$ for jet production as a function of $x_F$, at fixed pseudo-rapidity $\eta= 3.25$
and $\sqrt s$ = 500 GeV, compared with A$_N$DY data~\cite{Bland:2013pkt}. The central line is obtained adopting the best parameterisation of the Sivers functions compatible with SIDIS data and able to describe alone the inclusive pion production data. The shaded statistical error band is also shown.
\label{fig:an-andy-gamma}}
\end{figure}

In Fig.~\ref{fig:an-andy-gamma} we show our estimate, based on the chosen best parameterisation of the Sivers function (as obtained in the combined study of SIDIS and pion $A_N$ data, see the previous Section), for $A_N(x_F)$ in $\pup p\to {\rm jet}\, X$ processes at fixed pseudo-rapidity $\eta=3.25$ and $\sqrt s$ = 500 GeV, and compare it with the A$_N$DY data~\cite{Bland:2013pkt}. The corresponding statistical uncertainty (shaded area) is also shown.
The analogous estimates for $A_N(x_F)$ in $\pup p \to \gamma \, X$ processes at fixed pseudo-rapidity $\eta=3.5$ and $\sqrt s$ = 200 GeV and 500 GeV are given in Fig.~\ref{direct.g.P.S.}.

The very small values of $A_N$ for inclusive jet production at A$_N$DY are somehow consistent with the expectation within the GPM approach, where the behaviour of the Sivers functions from fits to the SIDIS data shows that the $u$ and $d$ quark Sivers functions have opposite sign but almost equal magnitude (at least in the valence region). Notice that in the estimates presented in Fig.~\ref{fig:an-andy-gamma} we adopt a parameterisation where the $u$-quark Sivers function at large $x$ is much larger than the $d$-quark function (this indeed allows to get a sizeable $A_N$ for inclusive $\pi^0$ production at forward rapidities (Fig.~\ref{fig:an-pion-Sivers})~\cite{Anselmino:2013rya}). Within the GPM approach the smaller size of $A_N$ in inclusive jet (at 500 GeV) w.r.t.~inclusive neutral pion (at 200 GeV) production at the same $x_F$ values is due to the different $p_T$ values covered (larger for A$_N$DY kinematics). With increasing $p_T$ the integration over the azimuthal phases becomes more suppressing, reducing the overall SSA.

Still concerning the SSA for inclusive jet production, a modified GPM with inclusion of initial and final state interactions~\cite{Gamberg:2010tj,D'Alesio:2011mc}, the so-called colour-gauge invariant (CGI) GPM approach, as well as the twist-3 approach~\cite{Gamberg:2013kla} would give a similar description of the data, but with an overall change of sign for the asymmetry. On the other hand the small values of $A_N$, due to strong cancellations between the $u$ and $d$ quark contributions, prevent any definite conclusion.

Results for single photon production are in some sense even more interesting. In such a case the quark charge factors lead to a dominance of the $u$-quark contribution and $A_N$ results sizeable. Once again, in the twist-3 approach~\cite{Kanazawa:2014nea} one gets a similar magnitude as in the GPM approach but with an opposite sign (see Fig.~\ref{direct.g.P.S.}). Thus, the measurement of $A_N$ for single photon production, no matter how difficult, would clearly discriminate between the two approaches.

\subsection{SSAs in $pp\to {\rm jet}\, h\,X$}
\label{pion-jet}

Azimuthal distributions of leading hadrons inside large transverse momentum jets in polarised proton-proton collisions are a very important and complementary source of information on TMD distribution and fragmentation functions.
In complete analogy with the SIDIS case, the process $p^\uparrow p\to {\rm jet}\,h\,X$ allows one to disentangle among the different TMD mechanisms at work (the Collins and Sivers effects notably) by looking at different angular moments of the observed azimuthal distribution.
The comparison with analogous SIDIS results, then,  offers a  unique way to check the universality and process dependence of TMDs.

{}From the experimental point of view, it is certainly not an easy task to measure the distribution of leading hadrons (mostly pions) around the reconstructed jet axis. However, as we have seen in Sec.~\ref{sec:1.exp.res}, work on these observables is in advanced progress at RHIC and first interesting preliminary results are already available, see Figs.~\ref{Collins.IFF}(a), \ref{Collins.FMS}, \ref{Collins.lG}.

{}From the theoretical side, the Collins asymmetry in the azimuthal distribution of leading pions  inside jets with large transverse momentum and (pseudo)rapidity in polarised $pp$ collisions was first studied in Refs.~\cite{Yuan:2007nd,Yuan:2008yv}.
By neglecting intrinsic motion in the initial protons, factorisation was proven for the TMD approach, opening the way for a direct check of the predicted Collins function universality by comparison with the corresponding Collins asymmetry measured in SIDIS processes. Notice that in this scheme, only the Collins azimuthal asymmetry is present at leading twist.

\begin{figure}[t!]
\hspace*{-0.1cm} \includegraphics[angle=0,width=0.5\textwidth]{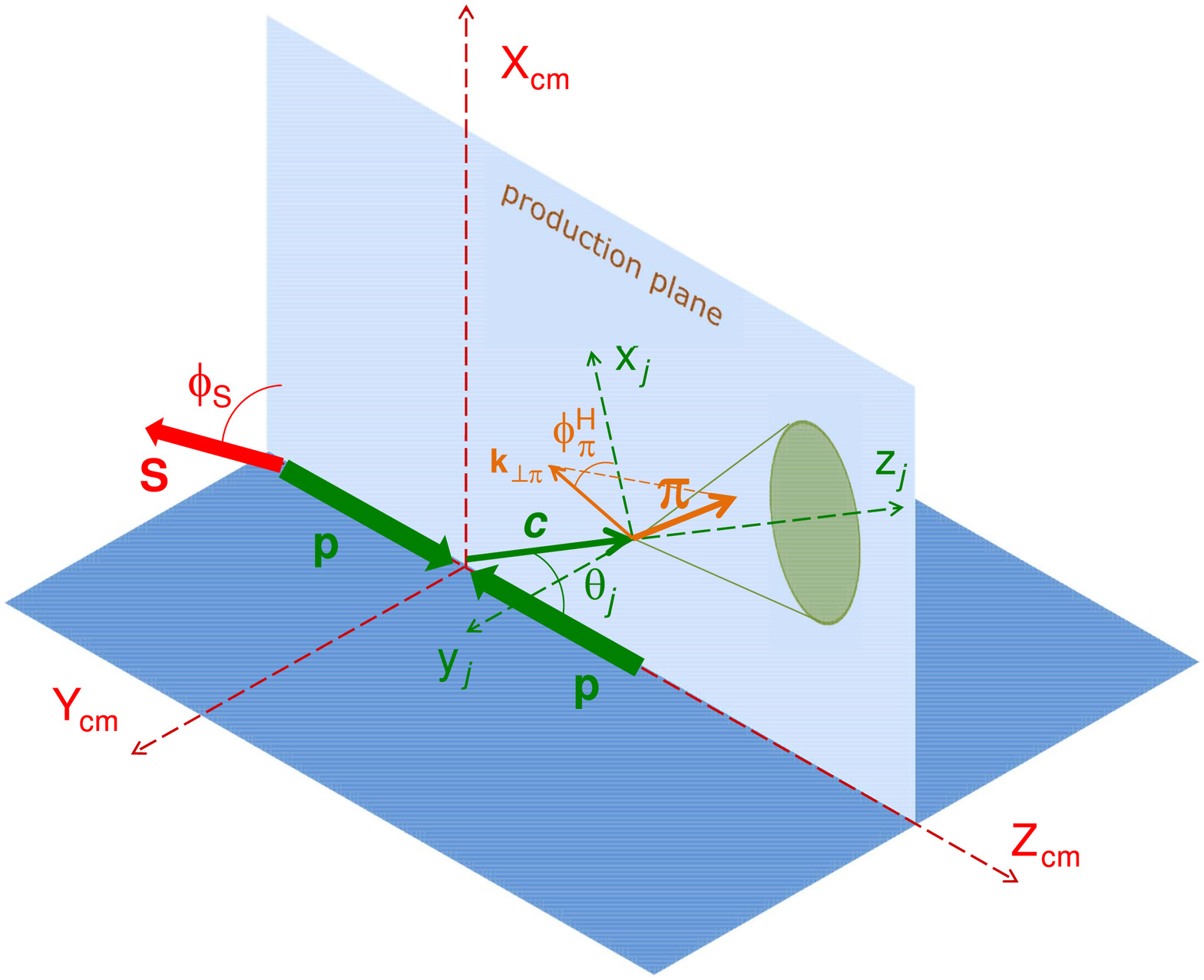}
 \caption{Kinematics for the process $p(S)\,p\to {\rm jet}\,  \pi X$ in the c.m.~frame of the two colliding protons.
 \label{fig:kinem} }
\end{figure}

In Refs.~\cite{D'Alesio:2010am,D'Alesio:2011mc,D'Alesio:2013jka} this approach was phenomenologically extended to include intrinsic motion effects also in the initial colliding protons, in the context of the generalised parton model. This makes the phenomenology much richer. In fact, besides the Collins asymmetry, also the Sivers asymmetry and analogous effects for gluons can be present and measured in principle. There are several other interesting aspects that the study of these azimuthal asymmetries can help clarifying: First of all, the unambiguous measurement of a non-vanishing Sivers azimuthal asymmetry in this process would be a clear indication that intrinsic motion inside initial colliding hadrons plays a significant role in the process and cannot be neglected; the measurement of azimuthal asymmetries generated by gluonic channels would offer the opportunity to study (un)polarised gluon TMDs at leading order, which is outside the reach of similar studies in SIDIS, DY processes or $e^+e^-$ collisions.
Moreover, at least in principle, this would also give a way to distinguish among quark and gluon generated jets.
As we have already seen, the lack of a factorisation proof for inclusive hadron production in $pp$ collisions
asks for a careful study of colour initial and final state interactions resulting in possible process dependences and universality breaking effects for TMDs (notice however that the Collins function is expected to be universal).
In Ref.~\cite{D'Alesio:2011mc} a first attempt has been made to incorporate in the generalised parton model initial and final
state interactions for quark initiated processes, in particular for the $p^\uparrow p\to {\rm jet}\, \pi\, X$ process.
The extension of this approach, known as the colour-gauge invariant generalised parton model (CGI-GPM),
to gluon channels and to other processes is under current investigation.
Furthermore, phenomenological applications incorporating all proper kinematical cuts for RHIC experiments, in particular jet cuts, are in progress~\cite{D'Alesio:2016}.
In this review we will limit ourselves to show a few representative results for the Collins and Sivers asymmetries, discussing their implications for future measurements.
We refer to the literature~\cite{D'Alesio:2010am,D'Alesio:2013jka} for all technical details and for a complete formulation of the approach and a discussion of the kinematics (see also Fig.~\ref{fig:kinem}).

\begin{figure*}[t!]
\resizebox{0.95\textwidth}{!}{%
 \includegraphics{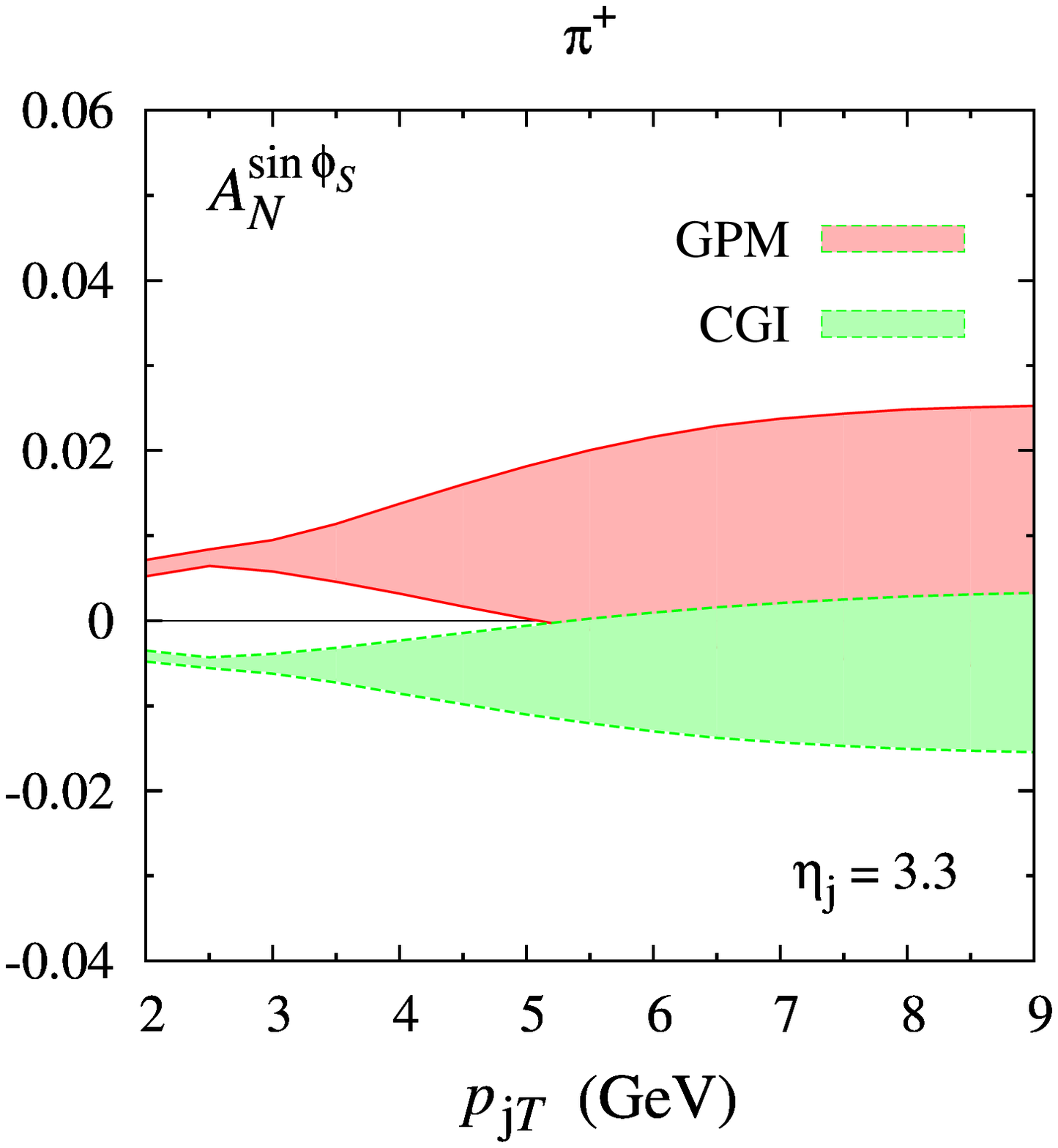}
 \includegraphics{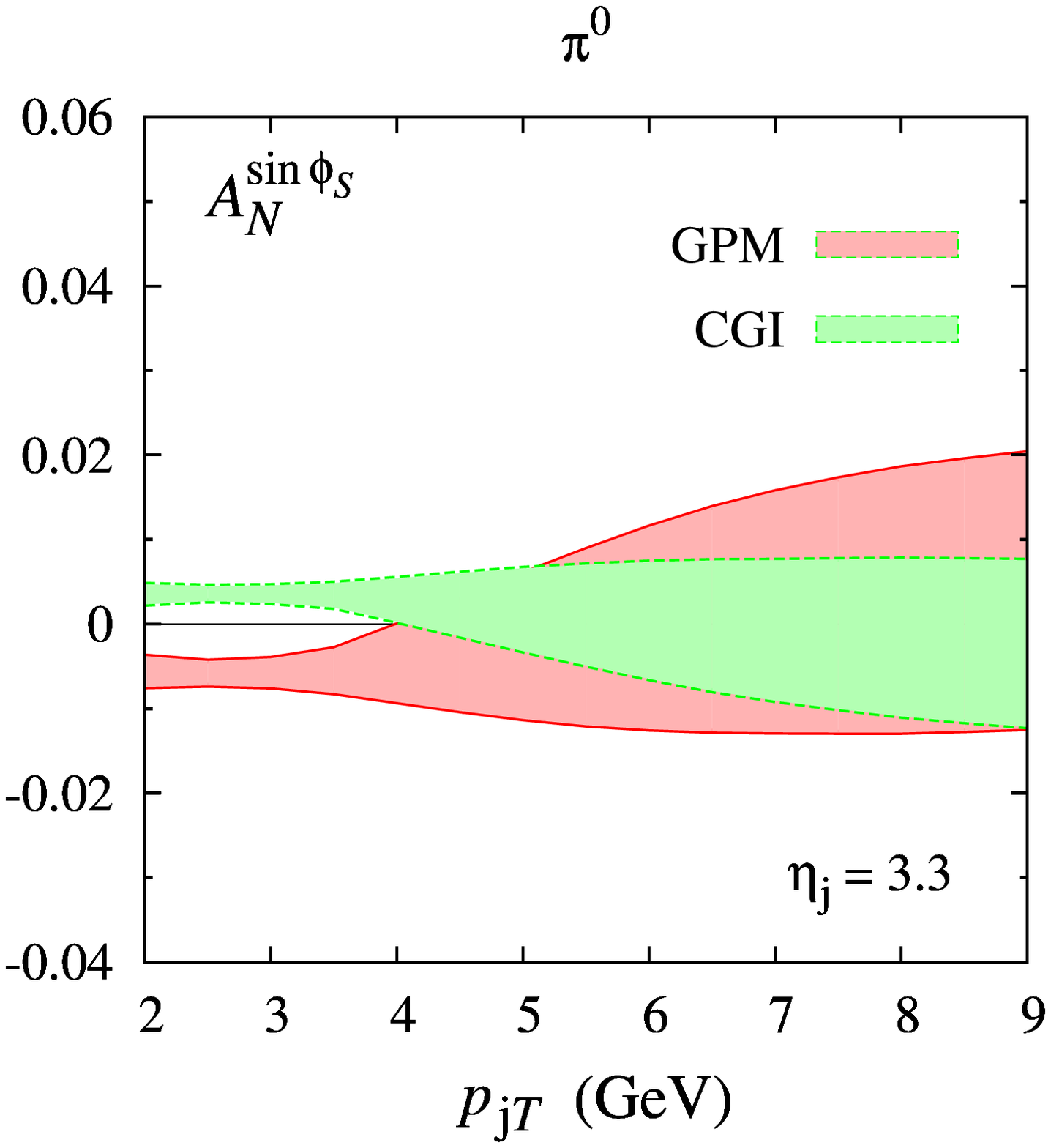}
 \includegraphics{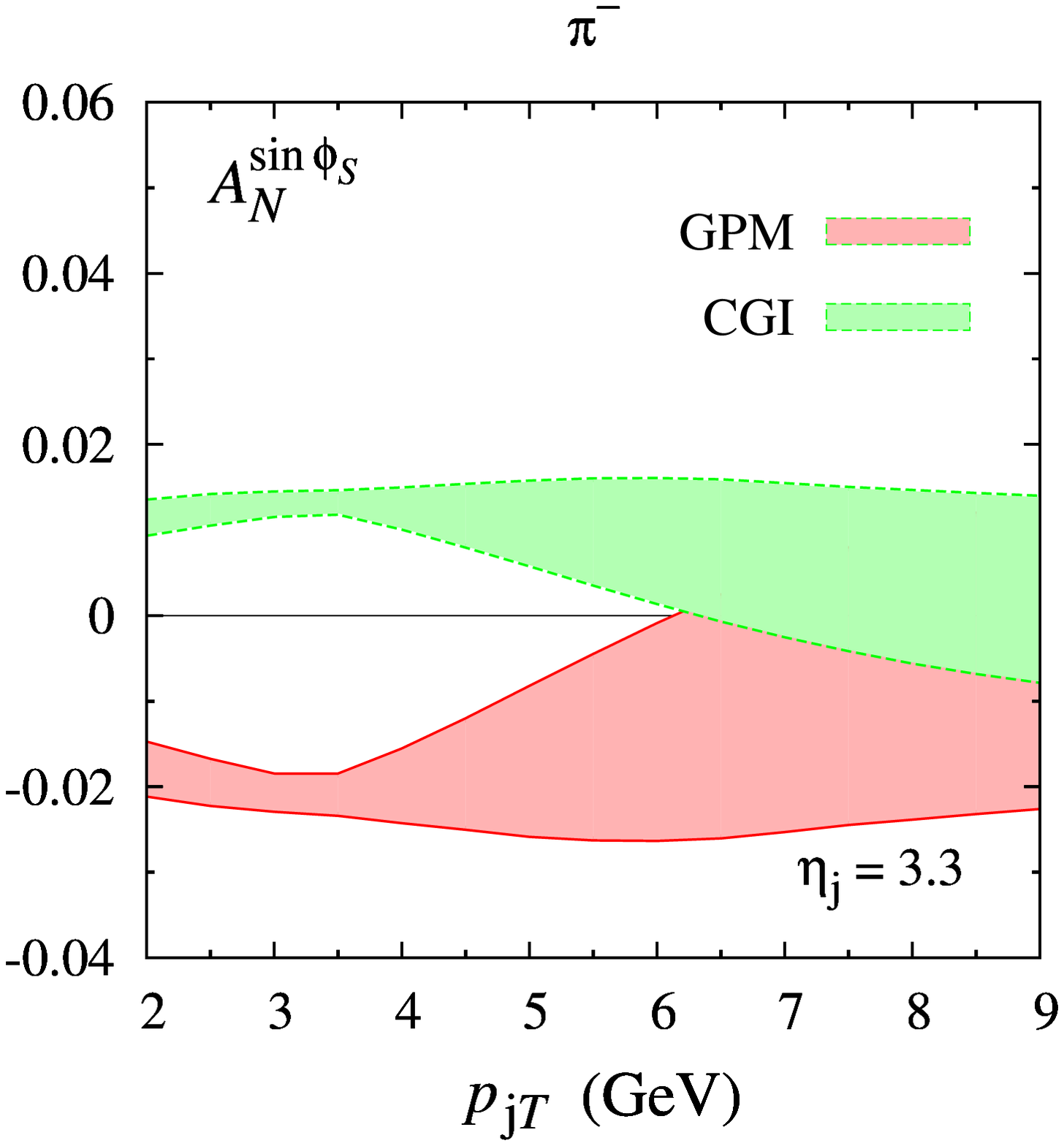}
}
\caption{Scan bands for the quark contribution to the Sivers asymmetry $A_N^{\sin\phi_{S}}$  in the GPM and CGI~GPM approaches, for the process $p^\uparrow \, p\to {\rm jet}\, \pi\,X$, as a function of $p_{{\rm j} T}$, at fixed value of the rapidity $\eta_{\rm j}$ and  c.m.~energy  $\sqrt{s}= 500$ GeV.
The shaded bands are generated following the scan procedure as detailed in Refs.~\cite{Anselmino:2012rq,Anselmino:2013rya}.}
\label{fig:siv-pijet}
\end{figure*}

The cross sections and the azimuthal asymmetries measured are given as a function of the following main kinematical variables: the total energy in the c.m.~frame of the colliding hadrons ($\sqrt{s} = 200$ and $500$ GeV for RHIC results);
the transverse momentum and the (pseudo)rapidity of the jet, $p_{{\rm j}T}$ and $\eta_{\rm j}$ respectively;
the light-cone momentum fraction of the observed hadron $h$, $z$; the hadron transverse momentum w.r.t.
the jet thrust axis (coinciding, in a leading order approach, with the fragmenting parton direction of motion), $\bm{k}_{\perp h}$; the azimuthal angles of the transverse spin of the initial polarised proton in the hadronic c.m. frame and of the observed hadron around the jet direction in the jet helicity frame, respectively $\phi_S$ and $\phi^H_h$.

As shown in Ref.~\cite{D'Alesio:2010am}, and considering the pion production case to be definite, the single transverse polarised differential cross section for the process $p(S) p\to {\rm jet}\, \pi\, X$ has the following leading-twist general structure [here $d\sigma$ is a shortand notation for $E_{\rm j}d\sigma/(d^3{\bm{p}}_{\rm j}\,dz\,d^2{ \bm{k}}_{\perp\,\pi})$]:
\begin{eqnarray}
2\,{\rm d}\sigma(\phi_{S},\phi_\pi^H) &\sim&  {\rm d}\sigma_0
+{\rm d}\Delta\sigma_0\sin\phi_{S}+
{\rm d}\sigma_1\cos\phi_\pi^H \nonumber\\
&+&
{\rm d}\Delta\sigma_{1}^{-}\sin(\phi_{S}-\phi_\pi^H)+
{\rm d}\Delta\sigma_{1}^{+}\sin(\phi_{S}+\phi_\pi^H)\nonumber\\
&+&{\rm d}\sigma_2\cos2\phi_\pi^H
+{\rm d}\Delta\sigma_{2}^{-}\sin(\phi_{S}-2\phi_\pi^H)\nonumber\\
&+&
{\rm d}\Delta\sigma_{2}^{+}\sin(\phi_{S}+2\phi_\pi^H)\,.
\label{eq:d-sig-phi-S}
\end{eqnarray}

The numerator and the denominator of the transverse spin asymmetry are given respectively by:
\begin{eqnarray}
&&{\rm d}\sigma(\phi_{S},\phi_\pi^H)-
{\rm d}\sigma(\phi_{S}+\pi,\phi_\pi^H)
\sim {\rm d}\Delta\sigma_0\sin\phi_{S} \nonumber\\
&+&
{\rm d}\Delta\sigma_{1}^{-}\sin(\phi_{S}-\phi_\pi^H)+
{\rm d}\Delta\sigma_{1}^{+}\sin(\phi_{S}+\phi_\pi^H)\nonumber\\
&+&{\rm d}\Delta\sigma_{2}^{-}\sin(\phi_{S}-2\phi_\pi^H)+
{\rm d}\Delta\sigma_{2}^{+}\sin(\phi_{S}+2\phi_\pi^H)\,,
\label{eq:num-asy-gen}
\end{eqnarray}
\begin{eqnarray}
&&{\rm d}\sigma(\phi_{S},\phi_\pi^H)+
{\rm d}\sigma(\phi_{S}+\pi,\phi_\pi^H)
 \equiv 2{\rm d}\sigma^{\rm unp}(\phi_\pi^H) \nonumber\\
&\sim&
{\rm d}\sigma_0 + {\rm d}\sigma_1\cos\phi_\pi^H+
{\rm d}\sigma_2\cos2\phi_\pi^H\,.
\label{eq:den-asy-gen}
\end{eqnarray}

In close analogy with the SIDIS case (see \textit{e.g.}~Ref.~\cite{Boglione:2015zyc}, this Special Issue), for the single spin asymmetry we can define appropriate azimuthal moments:
\begin{eqnarray}
&\!\!\!&A_N^{W(\phi_{S},\phi_\pi^H)}(\bm{p}_{\rm j},z,k_{\perp\pi})
\equiv
2\langle\,W(\phi_{S},\phi_\pi^H)\,\rangle(\bm{p}_{\rm j},z,k_{\perp\pi})\nonumber\\
&=&
2\,\frac{\int{\rm d}\phi_{S}{\rm d}\phi_\pi^H\,
W(\phi_{S},\phi_\pi^H)\,[{\rm d}\sigma(\phi_{S},\phi_\pi^H)-
{\rm d}\sigma(\phi_{S}+\pi,\phi_\pi^H)]}
{\int{\rm d}\phi_{S}{\rm d}\phi_\pi^H\,
[{\rm d}\sigma(\phi_{S},\phi_\pi^H)+
{\rm d}\sigma(\phi_{S}+\pi,\phi_\pi^H)]}\nonumber\\
&& \label{eq:gen-mom}
\end{eqnarray}
where $W(\phi_{S},\phi_\pi^H)$ is the suitable circular function required to single out the azimuthal dependent term of interest in the asymmetry.

For a full discussion of the azimuthal structure of the cross section and the asymmetry, and of the partonic contributions and the corresponding TMD PDFs and FFs involved in each term, see Ref.~\cite{D'Alesio:2010am}.
Here we will limit to discuss the terms that,
at least in principle, can be sizeable and are then phenomenologically relevant:\\
1) The Sivers asymmetry, $A_N^{\sin\phi_S}$, receiving contributions from both the quark and gluon Sivers functions in the transversely polarised proton, which cannot be disentangled;\\
2) The Collins asymmetry, $A_N^{\sin(\phi_S-\phi_\pi^H)}$, involving the quark transversity distribution in the
polarised proton and the quark Collins fragmentation functions;\\
3) The Collins-like asymmetry $A_N^{\sin(\phi_S-2\phi_\pi^H)}$,
involving linearly polarised gluons both inside the polarised proton and in the fragmentation process.

Due to the lack of space, we will show a few pictures just to give an idea of the type of results we may expect for kinematical configurations similar to those currently under investigation at RHIC. A more extensive and detailed treatment can be found in Refs.~\cite{D'Alesio:2010am,D'Alesio:2013jka}. The role of specific kinematical cuts on the variables of the process can be relevant. Theoretical studies specifically devoted to reproduce the RHIC kinematical configurations are in progress~\cite{D'Alesio:2016}.

Concerning the first two cases, Sivers and Collins effects, we will show results based on the scan procedure already described in the previous Section. This indeed represents, at the moment, the best way to study simultaneously the SSAs in SIDIS and $pp$ collisions.

In Fig.~\ref{fig:siv-pijet} we show the scan bands (envelope of all possible parameterisations) in the generalised parton model (red bands) and in the CGI-GPM (green bands) for charged and neutral pions in the $p^\uparrow p \to {\rm jet}\,\pi\,X$ process for the Sivers asymmetry $A_N^{\sin\phi_S}$, as a function of the jet transverse momentum, $p_{{\rm j}T}$, at fixed jet pseudorapidity, $\eta_{\rm j}=3.3$ and c.m.~energy $\sqrt{s}=500$ GeV. These estimates are obtained by adopting the quark Sivers function parameterisations resulting from the scan procedure in fitting the analogous Sivers asymmetry in SIDIS processes with the Kretzer FF set~\cite{Kretzer:2000yf}. Notice that at such forward pseudorapidities the gluon Sivers contribution is negligible. 

The uncertainty bands (corresponding to the shaded areas in the plots) reflect the indeterminacy in the quark Sivers function parameterisations due to the limited $x_B$ range explored by SIDIS experiments, $0.01 \leq x_B \leq 0.3$~\cite{Anselmino:2013rya}. Since the minimum allowed value of the light-cone momentum fraction for the quark inside the transversely polarised proton increases with the growing of $p_{{\rm j}T}$, the widening of the bands reflects the fact that for $p_{{\rm j}T} \geq 5.7$ GeV (corresponding to $x_F \geq 0.3$), the Sivers function parameterisations are poorly constrained.

{}From these results, we see that, with the support of precise data, in the important region $2\leq p_{{\rm j}T}\leq 4$ GeV one could be able to discriminate between the GPM and the CGI-GPM approach and test the process dependence of the Sivers function. For larger $p_{{\rm j}T}$ values the forthcoming SIDIS data at larger $x_B$ by JLab, constraining the large $x$ region of the Sivers function, will be very useful.
Notice also that the results for the neutral pion case are very similar to the analogous results for the Sivers asymmetry in the inclusive jet production, for which experimental data are available. At present, however, the width of the uncertainty bands does not allow to discriminate among the GPM and its colour gauge invariant version.

\begin{figure}
\begin{center}
\resizebox{0.30\textwidth}{!}{%
 \includegraphics{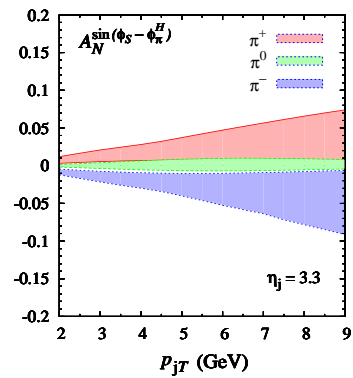}
}
\end{center}
\caption{Scan bands for the Collins azimuthal asymmetry $A_N^{\sin(\phi_{S}-\phi_\pi^H)}$  for the process $p^\uparrow \, p\to {\rm jet}\,
 \pi \, X$, as a function of $p_{{\rm j} T}$, at fixed value of the pseudorapidity,   $\eta_{\rm j}=3.3$ and c.m.~energy $\sqrt{s}= 500$ GeV.}
\label{fig:col-pijet}
\end{figure}

In Fig.~\ref{fig:col-pijet} we show results for the Collins azimuthal asymmetry for charged and neutral pions, $A_N^{\sin(\phi_S-\phi_\pi^H)}$, in similar kinematical configurations. These are obtained by adopting the parameterisations for the transversity distribution and the pion Collins function given by the scan procedure in the fit to the SIDIS Collins asymmetry data and to pion pair azimuthal correlations in $e^+e^-$ annihilations~\cite{Anselmino:2012rq}, using the Kretzer FF set. The meaning of the shaded areas is the same as discussed above. Comparison of these results with corresponding RHIC data can be very important in order to confirm the expected universality of the Collins function.

\begin{figure}
\begin{center}
\resizebox{0.50\textwidth}{!}{%
 \hspace*{-54pt}
 \includegraphics{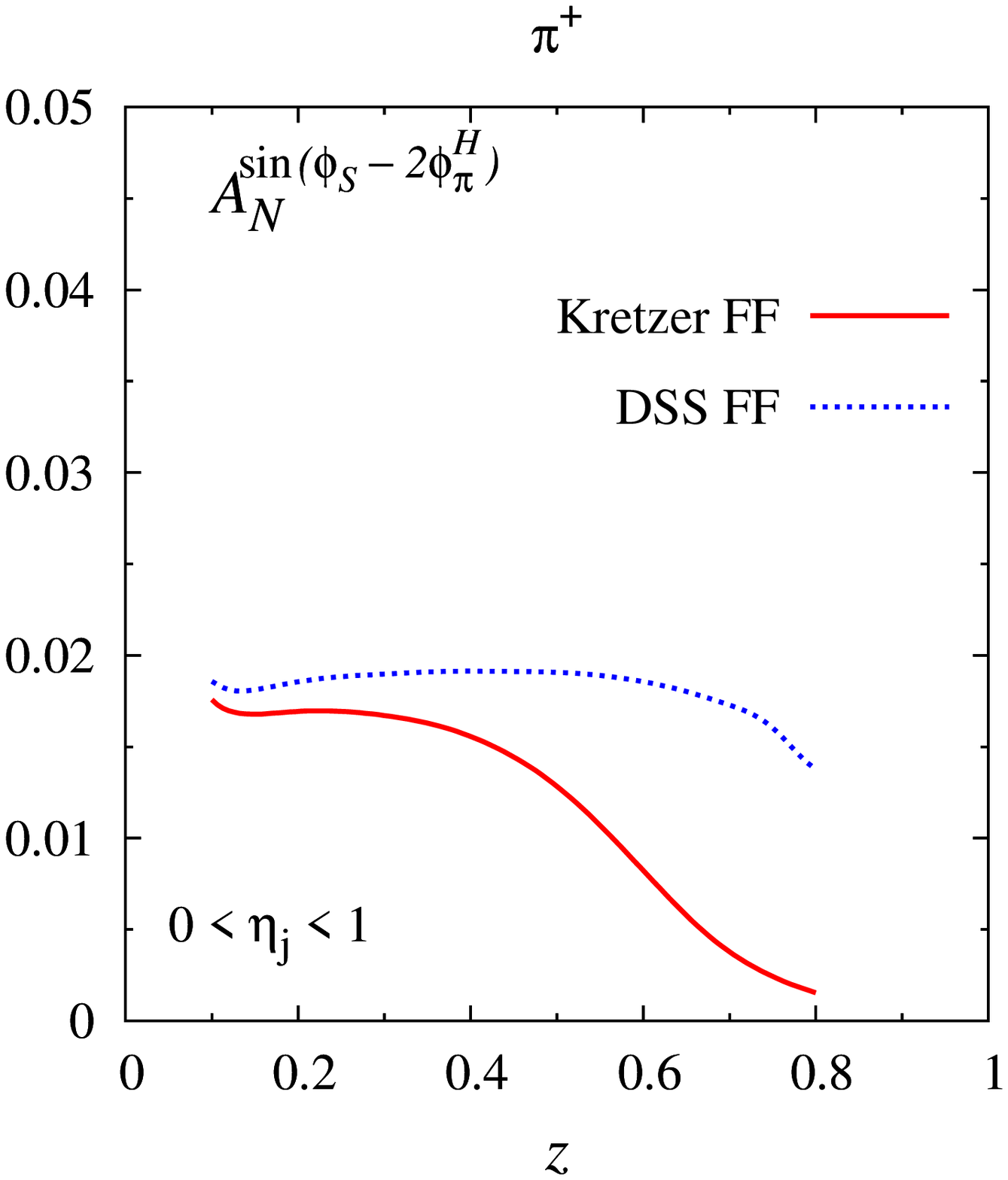}
 \hspace*{-54pt}
 \includegraphics{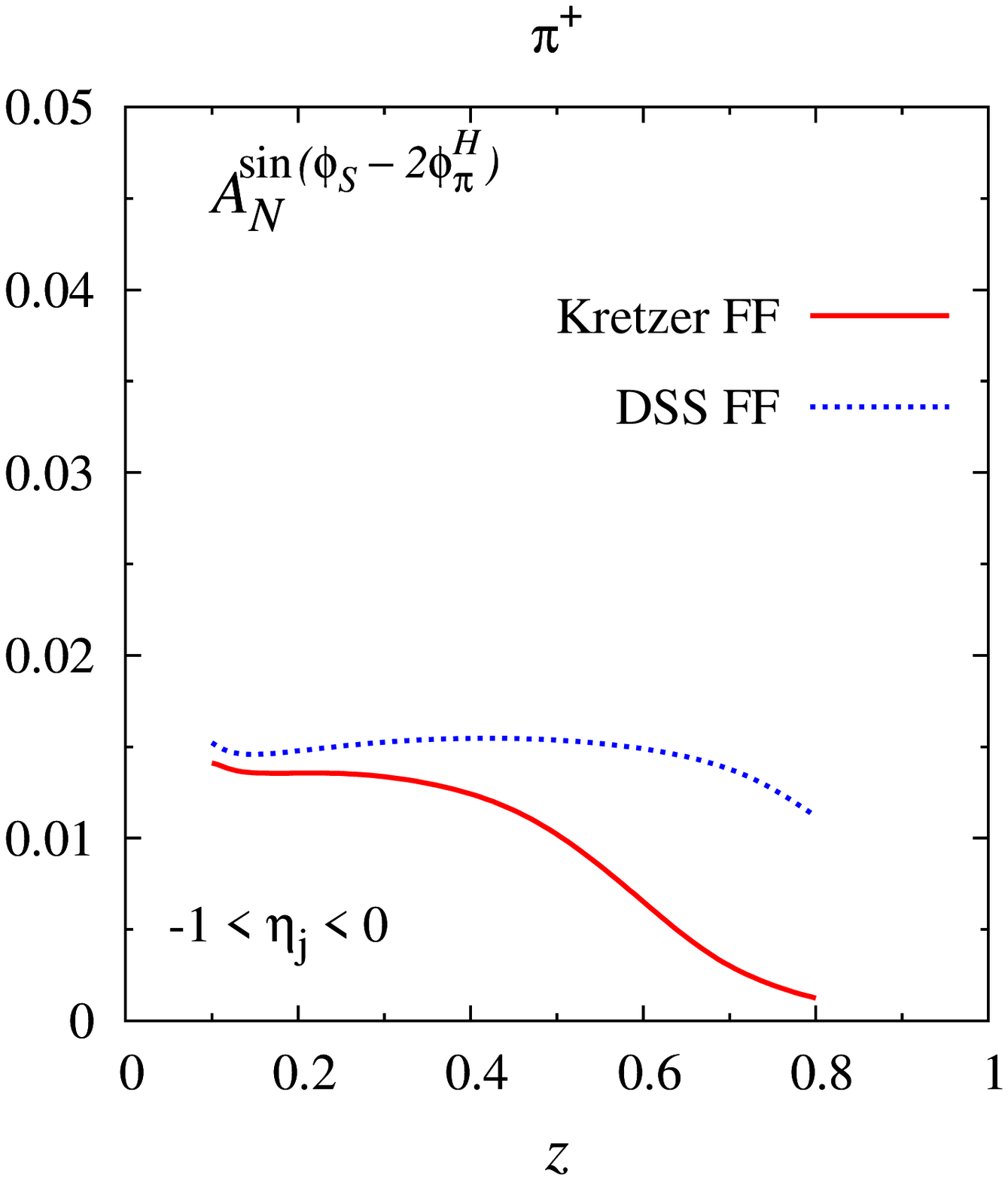}
}
\end{center}
\caption{Maximised estimates for the gluon Collins-like asymmetry $A_N^{\sin(\phi_S-2\phi_\pi^H)}$ for positively charged pions as a function of $z$, in the jet pseudorapidity range $0<\eta_{\rm j}<1$ (left panel) and $-1<\eta_{\rm j}<0$ (right panel) at $\sqrt{s} = 500$ GeV, adopting the Kretzer~\cite{Kretzer:2000yf} and the DSS~\cite{deFlorian:2007aj}  set of FFs.}
\label{fig:like-pijet}
\end{figure}

In Fig.~\ref{fig:like-pijet} we finally show maximised estimates for the gluon Collins-like asymmetry $A_N^{\sin(\phi_S-2\phi_\pi^H)}$ for positively charged pions as a function of the light-cone momentum fraction $z$, in the jet pseudorapidity range $0<\eta_{\rm j}<1$ (left panel) and $-1<\eta_{\rm j}<0$ (right panel) at $\sqrt{s} = 500$ GeV, adopting both the Kretzer~\cite{Kretzer:2000yf} and the DSS~\cite{deFlorian:2007aj} set of fragmentation functions. Notice that, since the gluon Collins-like TMDs are practically unknown at present, the sign of the asymmetry is arbitrarily chosen to be positive and all TMDs involved are saturated to their positivity bounds in order to maximise the contribution.
With some caution, since our estimates do not include all proper kinematical cuts, we can compare these results with the corresponding ones presented in Fig.~\ref{Collins.lG} for charged pions. From the data, there are some indications that the Collins-like asymmetries might be negative for both $\pi^+$ and $\pi^-$, very small at lower $z$ and with some tendency to increase at larger $z$ values, although error bars become larger also. More precise analyses and larger statistics for the data are required before drawing any conclusion on the gluon Collins-like functions. On the other hand these preliminary data may help in constraining the TMDs for linearly polarised gluons.

\subsection{SSAs in $pp\to \ell^+\ell^-\, X$ and $pp\to \ell \nu\, X  $}
\label{DY}

Drell-Yan processes, due to the lack of any fragmentation mechanism, are definitely an invaluable tool to study TMDs in the distribution sector.
Even with unpolarised beams by studying proper modulations in the azimuthal distribution of the di-lepton pair in its c.m.~frame one could access another interesting T-odd TMD, the so-called Boer-Mulders (BM) function, that gives the distribution of transversely polarised quarks inside an unpolarised nucleon.  More precisely we refer here to the large $\cos(2\phi)$ azimuthal dependence that can arise, at leading order, from a convolution of two Boer-Mulders distributions, as originally discussed in Ref.~\cite{Boer:1999mm} and further studied in Refs.~\cite{Boer:2002ju,Bianconi:2004wu,Sissakian:2008th,Barone:2010gk,Lu:2009ip,Pasquini:2014ppa}. Unfortunately the extraction of this TMD is still affected by too many sources of uncertainty.

Moving to the SSAs, our main topic, DY represents definitely one of the cleaner processes to access the Sivers effect.
Indeed for such processes, like for SIDIS processes and in contrast to inclusive single hadron production, TMD factorisation has been proven to hold~\cite{Collins:1984kg,Ji:2004xq,Collins:2011zzd,GarciaEchevarria:2011rb}. Moreover, according to the widely accepted interpretation of the QCD origin of tSSAs as final or initial state interactions of the scattering partons~\cite{Brodsky:2002cx}, the Sivers function should exhibit opposite signs in SIDIS and DY processes~\cite{Collins:2002kn}. This property, still to be confirmed by experiments, represents one of the major challenges in our understanding of SSAs.

Predictions for Sivers $A_N$ in DY at different forthcoming or foreseen experimental setups were given in Ref.~\cite{Anselmino:2009st}, which we follow here. For their relevance, in the sequel we focus on RHIC and COMPASS kinematics. In the second part of this Section we will comment on the very interesting case of DY via $W$-boson exchange.

In Ref.~\cite{Anselmino:2009st} a detailed study of the SSAs for DY processes in the $\pup-p\,$ c.m.~frame was presented, by analysing their dependence on $q^2=M^2$, the square invariant mass of the lepton pair, representing the large scale in the process, and $q_T^2$, the small one, where $q$ is the lepton-pair four-momentum. In order to collect data at all azimuthal angles, one defines the following moment of the spin asymmetry:
\be
 A_N^{\sin(\phi_{\gamma} - \phi_S)} \equiv
\frac{\int_0^{2\pi} d\phi_{\gamma} \>
[d\sigma^{\uparrow} - d\sigma^{\downarrow}] \>
\sin(\phi_{\gamma}-\phi_S)}
{\frac{1}{2}\int_{0}^{2\pi} d\phi_{\gamma} \>
[d\sigma^{\uparrow} + d\sigma^{\downarrow}]} \,,\label{ANW}
\ee
where $\phi_\gamma$ and $\phi_S$ are respectively the azimuthal angle of the
$\ell^+\ell^-$ pair and of the proton transverse spin, and
\bea
d\sigma^{\uparrow} - d\sigma^{\downarrow} &=& \sum_q e_q^2 \int d^2\bfk_{\perp 1} \, d^2\bfk_{\perp 2}\,
\delta^2(\bfk_{\perp 1} + \bfk_{\perp 2} - \bfq_T) \nonumber \\
&\times & \frac{4\pi\alpha^2}{9M^2} \Delta^N\!\hat f_{q/\pup}(x_1, \bm{k}_{\perp 1}) \>
 f_{\bar q/p}(x_{2}, k_{\perp 2})
 \> . \label{ANW2}
 \eea
In order to present estimates for the Sivers asymmetries in Drell-Yan processes -- and test the crucially important sign change when going from SIDIS to DY -- we insert the Sivers functions as extracted from SIDIS fits~\cite{Anselmino:2008sga} into Eq.~(\ref{ANW2}), changing their sign.

\begin{figure*}
\centering
  \includegraphics[totalheight=0.27\textheight,angle=0]{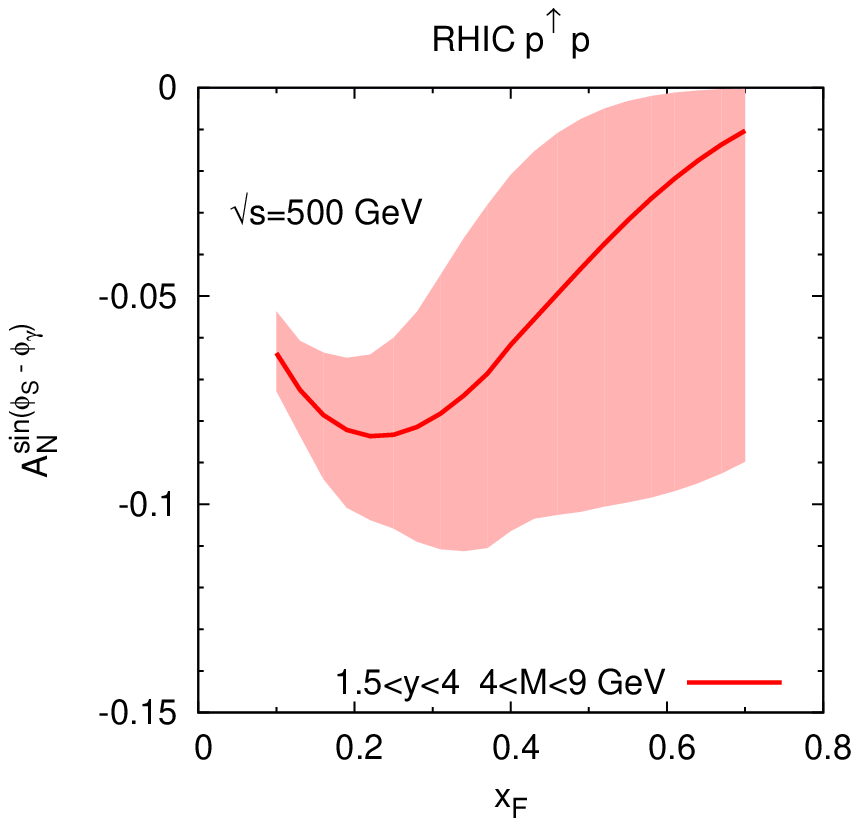}~\hspace*{-1.8cm}
  \includegraphics[totalheight=0.27\textheight,angle=0]{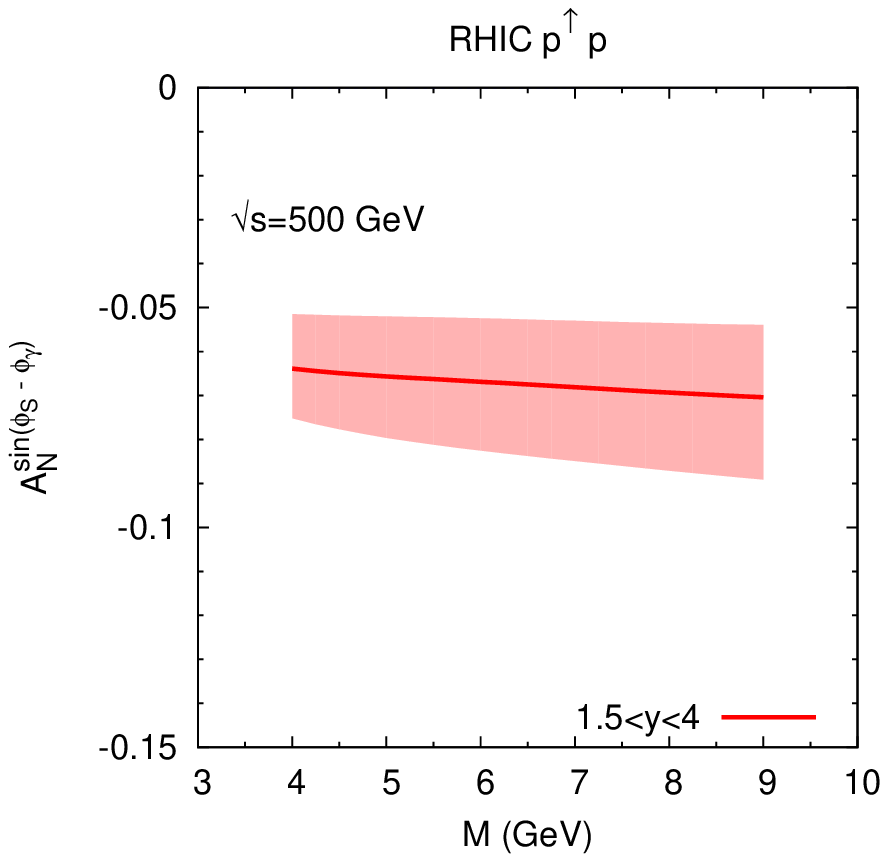}
\caption{$A_N^{\sin(\phi_{\gamma}-\phi_S)}$ for the Drell-Yan process $p^{\uparrow} p\to \mu^{+}\mu{^-}\,X$ at RHIC, as a function
of $x_F$ (left panel) and $M$ (right panel). See text for details.}
\label{fig:RHIC-AN}       
\end{figure*}

In Fig.~\ref{fig:RHIC-AN} we show $A_N^{\sin(\phi_{\gamma}-\phi_S)}$ for RHIC kinematics as a function of $x_F=x_1-x_2$ (at leading order) and $M$ at $\sqrt s = 500$~GeV, integrated over $q_T$ in the range $0 \leq q_T \leq 1$ GeV, which is within the region of validity of the TMD approach. The other integration ranges are $4 \leq M \leq 9$ GeV, at fixed $x_F$, for the left plot, and $0.1 \leq x_F \leq 0.7$, at fixed $M$, for the right plot, with the further constraint $1.5 \leq y \leq 4$, according to the experimental kinematical cuts. This region includes the range already explored by SIDIS measurements, where the Sivers functions are reliably constrained, extending it up to much higher values of $x$. The maximum value (in magnitude) of the asymmetry ($\sim$ 8\%) shown in the left panel is expected at $x_F\sim x_1 \simeq 0.2$, where the valence Sivers functions reach their maximum. The asymmetry for $x_F>0.4$, that implies $x$ values not covered by SIDIS data, is affected by a huge uncertainty band (calculated according to the procedure described in Ref.~\cite{Anselmino:2008sga}).
Measurements in this region could then offer an opportunity to test also the large-$x$ behaviour of the Sivers functions. Moreover, data in the negative $x_F$ region would test the contribution of the sea Sivers functions, as first pointed out in Ref.~\cite{Collins:2005rq}.

\begin{figure*}
  \centering
  \includegraphics[totalheight=0.4\textheight,angle=-90]{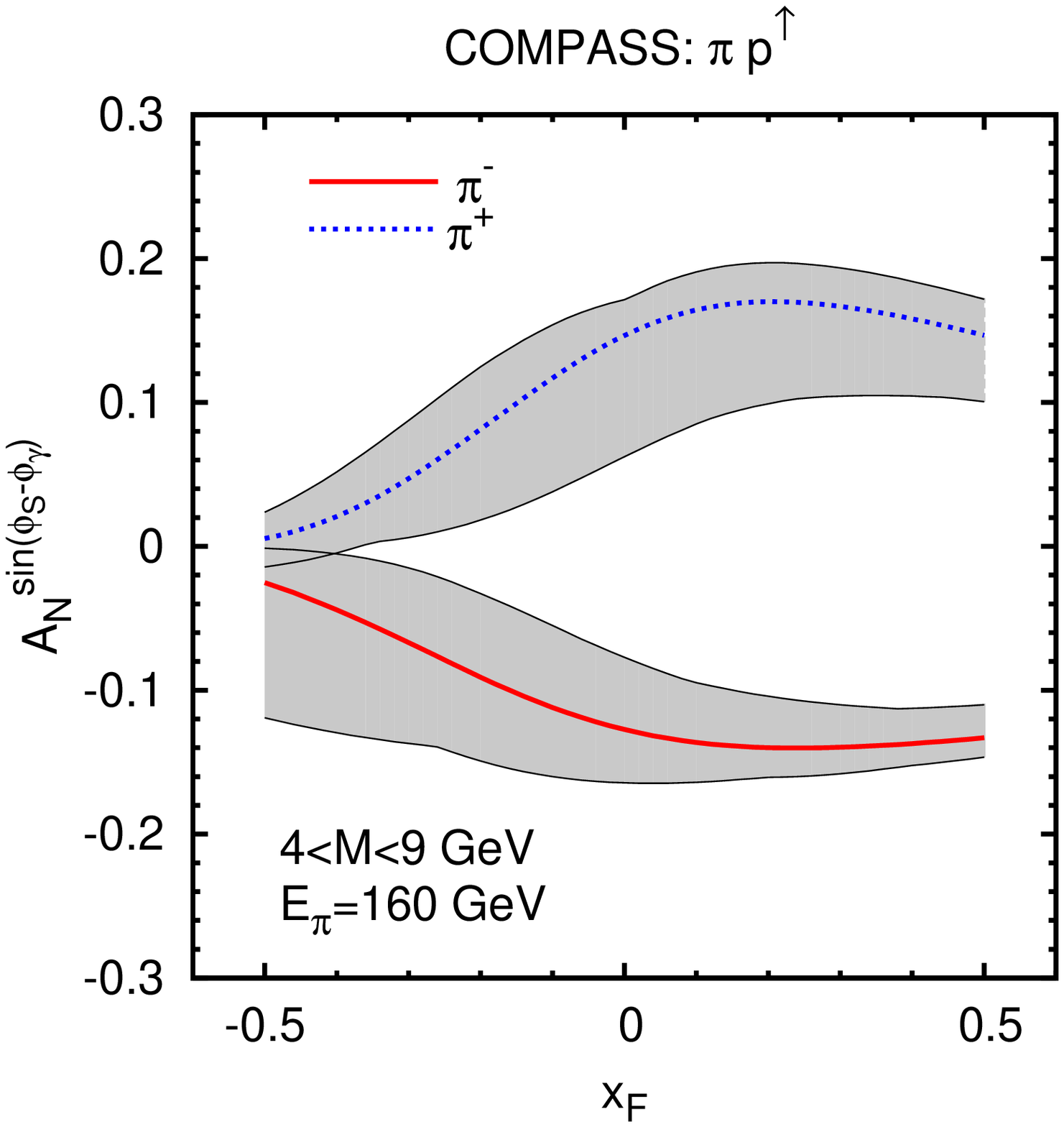}~\hspace*{-1.8cm}
  \includegraphics[totalheight=0.4\textheight,angle=-90]{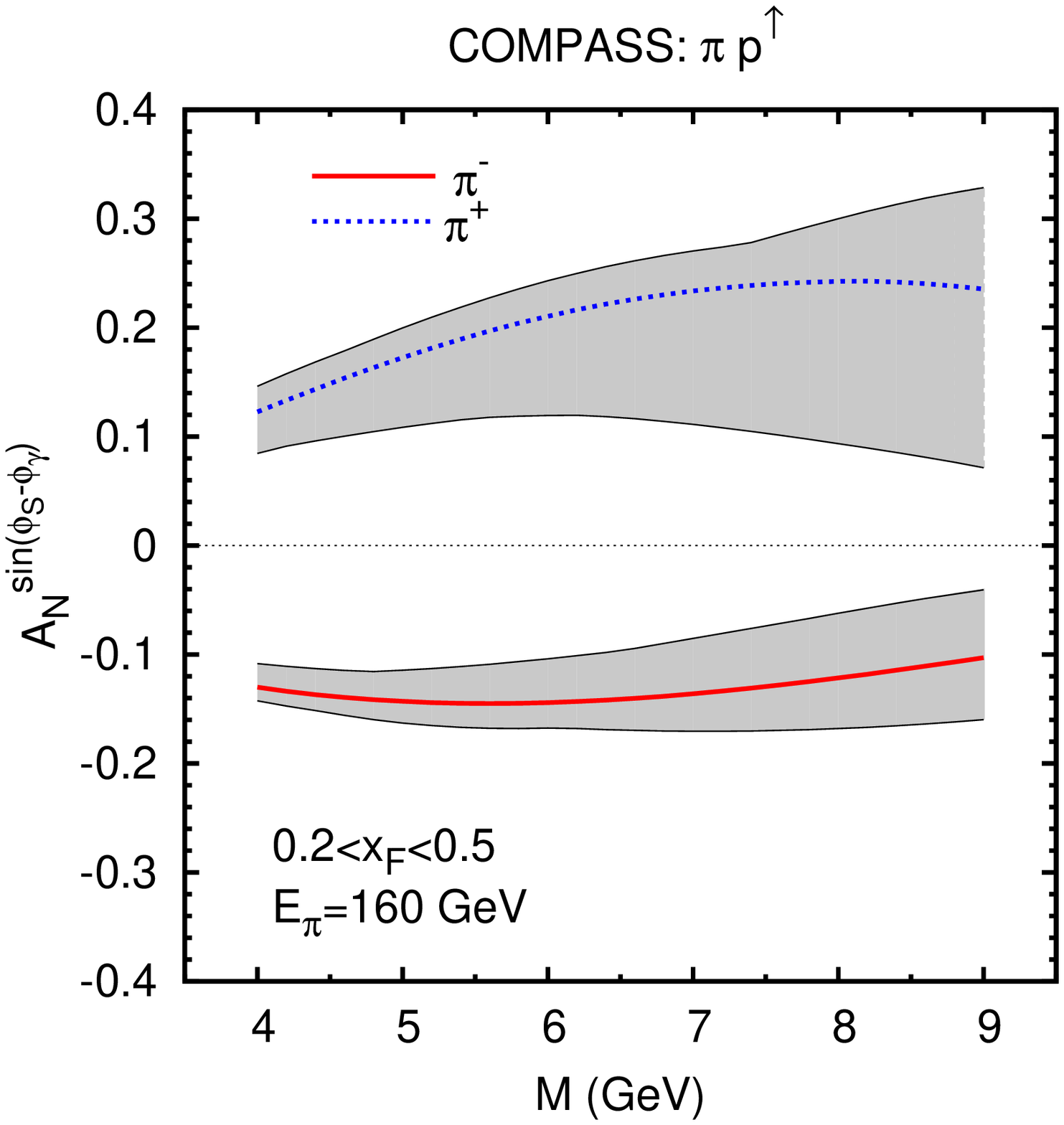}
\caption{$A_N^{\sin(\phi_{\gamma}-\phi_S)}$ for the Drell-Yan process $\pi^\pm p ^\uparrow \to \mu^{+}\mu^{-}\,X$ at COMPASS, as a
function of $x_F $ (left panel) and $M$ (right panel). See text for details.}
\label{fig:COMPASS-AN}       
\end{figure*}

In Fig.~\ref{fig:COMPASS-AN} results for $A_N^{\sin(\phi_{\gamma}-\phi_S)}$ at COMPASS are shown. In such a case the process under consideration is $\pi^\pm  p^\uparrow \to \mu^+\mu^- X$. Notice that, by rotational invariance, one has
\be
A_N^{A^\uparrow B \to \ell^+ \ell^- X}(x_F, \phi_\gamma) =
A_N^{B A^\uparrow \to \ell^+ \ell^- X}(-x_F, \phi_\gamma) \>.
\label{rel}
\ee

Again both the numerator and denominator of Eq.~(\ref{ANW}) are integrated over $q_T$ up to 1 GeV, $M$ between 4 and 9 GeV (left plot) and $x_F$ in the range 0.2-0.5 (right plot). The pion beam energy, in the laboratory frame, is taken to be 160 GeV, corresponding to $\sqrt s = 17.4$ GeV. It is interesting to notice that positive values of $x_F$ correspond to (or at least overlap with) the $x$ region explored by the SIDIS experiments ($x \leq 0.3$) and the data used to extract our Sivers functions. Instead, negative values of $x_F$ correspond to larger $x_2$ values. Finally, it is important to remark that, as the COMPASS experiment will involve charged pion beams, $\pi^-(\bar{u}d)$ and $\pi^+(u\bar{d})$, the dominant elementary process contributing to the asymmetry will be $\bar u _{\pi^-} u_p \to \mu^+\mu^-$ for the $\pi^-$ beam and $\bar d _{\pi^+} d_p\to \mu^+\mu^-$ for the $\pi^+$ beam. Consequently, the
prediction of the sign change of the Sivers function in SIDIS and Drell-Yan processes~\cite{Brodsky:2002rv,Collins:2002kn} can be clearly tested, as the
sign of the asymmetry for $\pi^-$ ($\pi^+$) beam is given by the sign of the $u$ ($d$) Sivers function, which is well established.

It is also worth to remark that in standard DY processes, that is via photon exchange, the up-quark flavour is enhanced w.r.t.~the down one through the electric charge factor. On the other hand the opposite sign of their corresponding Sivers functions could lead to a partial cancellation, reducing eventually the SSA. On top of that, there is only a moderate sensitivity to the sea-quark contributions. To have a better flavour separation, in Refs.~\cite{Brodsky:2002pr,Schmidt:2003wi} the authors proposed to measure $A_N$ for the analogous process $p^\uparrow p \to W \, X \to \ell \nu \, X$ at RHIC.
In such a case at forward rapidities w.r.t.~the polarised proton the $W^+$ production would be directly sensitive to the up-quark Sivers function, while the $W^-$ to the down one, leading to a well defined sign for $A_N$. This could play a crucial role for testing the time-reversal modified universality of the Sivers function.
At the same time, by looking at negative rapidities, one could also have a more direct access to the sea-quark Sivers distributions.
This idea was further developed in Ref.~\cite{Kang:2009bp}, where, to avoid the potential difficulties in reconstructing the $W$ boson (actually solved at RHIC), also the lepton production from the $W$ decay was considered. In that work it was shown how the SSA for inclusive $W$ production could be much larger than the corresponding $A_N$ in DY processes and that, although the lepton asymmetry is diluted from the $W$ decay, it is still at the level of several percents, with a clear signature and measurable for a good range of lepton rapidity at RHIC.

More recently the same calculation has been performed taking into account the QCD evolution of the Sivers function~\cite{Echevarria:2014xaa}, leading to much lower $A_N$ values, for the inclusive $W$ production, w.r.t.~those calculated in Ref.~\cite{Kang:2009bp}. On the other hand there is still some controversy and many uncertainties on some phenomenological aspects of the evolution of TMDs, in particular concerning the proper treatment of their nonperturbative part. For these reasons these results have to be taken with some caution and, as also recently pointed out in Ref.~\cite{Huang:2015vpy}, it cannot be excluded that TMD evolution could lead to a much lower overall effect (see Fig.~\ref{AN.W.Z0.Proj} and comments in Sec.~\ref{sec:3}). Notice that in Refs.~\cite{Kang:2009bp,Echevarria:2014xaa} they considered directly $A_N=(d\sigma^\uparrow-d\sigma^\downarrow)/(d\sigma^\uparrow+d\sigma^\downarrow)$ at fixed $\phi_\gamma = 0$, with the polarised proton moving along the $+z$ direction (and $\phi_S=\pi/2$). That means an opposite sign (but the same size) compared to the moment of the asymmetry, $A_N^{\sin(\phi_{\gamma}-\phi_S)}$.

\subsection{SSAs in double inclusive production in $pp$ collisions}

In Sec.~\ref{pion-jet} we have discussed in some detail the importance of the study of SSAs in $pp\to {\rm jet}\, \pi\,X$, where the two particles in the final state belong to the same hemisphere. Other interesting cases of double inclusive production in $pp$ collisions, with the two final particles belonging to opposite hemispheres, were proposed in Refs.~\cite{Boer:2003tx,Bacchetta:2005rm,Bomhof:2007su,Bacchetta:2007sz,Boer:2007nh}. We refer here to the study of di-jet, hadron-jet, photon-jet production at large $P_T$ in hadronic processes, where a second small scale, relevant for the TMD approach, is the total $q_T$ of the two particles in the final state (i.e.~their $q_T$ imbalance). Note that this $q_T$ is of the order of the intrinsic partonic momentum $k_\perp$. In close analogy to the CGI-GPM, this approach leads to a modified TMD factorisation scheme, with the inclusion in the elementary processes of colour gauge link factors~\cite{Bomhof:2004aw,Bomhof:2006dp,Ratcliffe:2007ye}. Despite the identification of two separate scales, even for such hadronic processes some problems with the TMD factorisation  have been pointed out~\cite{Rogers:2010dm,Rogers:2013zha}.

For their relevance in the context of TMD factorisation and process dependence, less involved processes from the point of view of colour gauge links need to be considered. For example, processes like $p^\uparrow p\to \gamma\gamma\,X$~\cite{Qiu:2011ai} or $p^\uparrow p\to J/\psi\; \gamma\,X$, where only initial state interactions are involved like in DY. These could play a significant role in shedding light on these issues.

\section{Conclusions and outlook}

Transverse single spin asymmetries in hadronic reactions, starting from the first observations carried out almost 40 years ago, still represent one of the most fascinating observables in hadron physics. Their size and features and their persisting at the current available c.m.~energies and $p_T$ values are puzzling and, at the same time, challenging aspects in QCD. The study of SSAs is also strongly related to our understanding of the structure of hadrons and their spin and orbital angular momentum content in terms of partons. In this sense they play a crucial role in the 3D mapping of the nucleons.

Certainly the start of RHIC and its transverse polarised proton program has provided till today a wealth of new data, which offer the unique opportunity to expand our current one-dimensional picture of the nucleon by imaging the proton in both momentum and impact parameter space. At the same time we can further understand the basics of colour interactions in QCD and how they manifest themselves in different processes.
The most recent results from PHENIX and STAR have shown that large transverse single spin asymmetries for inclusive hadron production, seen in $pp$ collisions at fixed-target energies and modest $p_T$, extend to the highest RHIC energies and surprisingly large $p_T$.

In this review we have focused in some detail on the GPM approach, that gives also the possibility to access directly the 3D structure of the nucleons. Even if not formally proven, this model is able to reproduce fairly well many features of the available data, and at the same time, represents a window on possible factorisation breaking effects, not yet seen. We have presented and discussed a selection of relevant experimental results from RHIC and given estimates for several inclusive processes within the GPM approach, pointing out the similarities and the differences with respect to the twist-3 results. In particular, it has been shown how, among the single inclusive processes, tSSAs in direct photon production could represent a clean observable to access the mechanism at work in the initial state, allowing to disentangle these two approaches. Midrapidity data in inclusive pion or jet production have been shown to provide a powerful tool to constrain the still poorly known gluon Sivers function.

However, SSAs in inclusive processes present some features that deserve more attention and study. Indeed there are two surprising facts, which might indicate that the underlying subprocess causing a significant fraction of the large transverse single spin asymmetries in the forward direction for single inclusive production are not dominated by  2 $\rightarrow$ 2 parton scattering processes; first: $A_N$ at forward rapidities basically vanishes going from inclusive $\pi^0$s to electromagnetic jets and second:  the asymmetries are basically flat as a function of $p_T$.  During the 2015 transversely polarised $pp$ run at $\sqrt{s}= 200$ GeV the conjecture that the subprocesses dominating the forward $A_N$ are of diffractive nature can be definitely tested by STAR, by measuring $A_N(\pi^0)$ for single and double diffractive events and tagging one or both forward scattered protons in the STAR Roman Pots.

The above issues questioning the single inclusive processes could be clarified by studying SSAs in double inclusive production, proposed in recent years as a tool to access separately the contributions from the initial and final state effects. The results from transversely polarised data taken in 2006, 2011, and 2012, for example, demonstrate for the first time that transversity is accessible in polarised proton collisions at RHIC at $\sqrt{s}$ = 200 GeV and 500 GeV through double inclusive observables involving the Collins FFs or the IFFs.
In particular, we have discussed SSAs for pion-jet production, where, at variance with the single inclusive case, one could disentangle the Sivers and the Collins effects and test their universality properties. In this context a surprising result is the fact that the measurements exhibit basically no dependence on $\sqrt{s}$, despite the jet-$p_T$ differs by a factor 2. These data could then provide valuable inputs to constrain TMD evolution effects. Moreover, SSAs in pion-jet production allow to access, and constrain, the linearly polarised gluon distributions both in the initial and final state.

From the theoretical point of view, while the Collins function is proven to be universal, according to the present understanding of TMDs in QCD, the Sivers function is expected to be process dependent, changing its sign when probed in DY w.r.t.~SIDIS processes.  Estimates for SSAs in Drell-Yan processes, that could provide a clear-cut proof of this modified universality, have also been given. The ongoing program at COMPASS on tSSAs for DY processes with charged pion beams would be an invaluable opportunity in this respect. The first experimental investigation of this non-universality has been already provided by STAR, measuring $A_N$ for $W^{\pm}$ bosons.
High precision data for $A_N$ for $W^{\pm}, Z^0$  boson, DY production, at $\sqrt{s} =  500$ GeV, becoming available after the transversely polarised $pp$ RHIC Run in 2017, will provide a unique opportunity for the ultimate test of the theoretical concept of TMDs, factorisation, evolution and non-universality. The measurement of $A_N$ for $W^{\pm}$ bosons as a function of rapidity will also give the worldwide first constraint on the light sea-quark Sivers functions.

Among the most relevant open issues we have to point out that TMD evolution still deserves further studies on the theory side and new SSA measurements in $pp$ collisions could definitely help in clarifying its effective role. Another point to be investigated is whether the underlying mechanisms responsible for the tSSAs observed in inclusive $pp$ collisions and SIDIS processes have or not the same origin. A true global analysis, still missing, could allow to check this very important aspect. In this respect the forthcoming experimental programs like, for example, those planned or proposed at the Electron Ion Collider, AFTER@LHC and SeaQuest at FermiLab would be extremely helpful.

The rich experimental activity in progress and the current developments and refinements in the theoretical approaches are a guaranty that transverse single spin asymmetries will keep providing novel and deep insights into the 3D structure of the nucleons.

\acknowledgement
E.C.A.~is grateful to her colleagues in the spin working groups of STAR and PHENIX. This work was supported by the U.S.~Department of Energy under Contract No.~DE-SC0012704. U.D. and F.M.~acknowledge the long and fruitful collaboration with M.~An\-selmino, M.~Boglione, E.~Leader, S.~Melis, C.~Pisano and A.~Pro\-kudin.

\end{document}